\documentclass[9pt,twocolumn]{extarticle}

\usepackage[margin=2cm]{geometry}
\usepackage{authblk}
\usepackage{graphicx}
\usepackage{float}
\usepackage{amsmath}
 \usepackage{booktabs}
\usepackage[square,numbers,sort&compress]{natbib}
\usepackage{enumitem}
\usepackage{hyperref}
\usepackage[switch]{lineno}

\usepackage{relsize}
 \usepackage[T1]{fontenc}
 
\usepackage{siunitx}
  \newcommand{\ab}{\alpha \textrm{-} \beta}

  \newcommand{\ie}{\textit{i.e.}}
  \newcommand{\eg}{\textit{e.g.}}
  \newcommand{\ktrans}{k_\mathrm{trans}}
  \newcommand{\kfold}{k_\mathrm{fold}}
  
  \newcommand{\ttrans}{\tau_{\textrm{trans}}}
  \newcommand{\tfold}{\tau_{\textrm{fold}}}
  \newcommand{\tmin}{\tau_{\textrm{min}}}
  \newcommand{\tmax}{\tau_{\textrm{max}}}  
  
  \newcommand{\tonset}{\tau_{\textrm{onset}}}  
    
  \newcommand{\asyma}{\textrm{asym}_{\alpha}}
  \newcommand{\asymb}{\textrm{asym}_{\beta}}

\usepackage{xcolor}
\definecolor{YKB}{rgb}{0.00,0.18,0.65}

\title{\textbf{Structural asymmetry along protein sequences\\
and co-translational folding}}

  \author[1,*]{John M. McBride}
  \author[1,2,*]{Tsvi Tlusty}

  \affil[1]{Center for Soft and Living Matter, Institute for Basic Science, Ulsan 44919, South Korea}
  \affil[2]{Departments of Physics and Chemistry, Ulsan National Institute of Science and Technology, Ulsan 44919, South Korea}
  \affil[*]{jmmcbride@protonmail.com, tsvitlusty@gmail.com}

\begin{document}

\maketitle

\section*{Abstract}
\textbf{
  Proteins are translated from the N- to the C-terminus, 
  raising the basic question of how 
  this innate directionality affects their evolution.
  To explore this question, we analyze \num{16200} structures from 
  the protein data bank (PDB). 
  We find remarkable enrichment of $\alpha$-helices
  at the C terminus and $\beta$-strands at the N terminus. 
  Furthermore, this $\ab$~asymmetry correlates with sequence length and contact order,
  both determinants of folding rate, hinting at possible links to co-translational folding (CTF).
  Hence, we propose the `slowest-first' scheme, whereby 
  protein sequences evolved structural asymmetry to accelerate CTF: 
  the slowest of the cooperatively-folding segments are positioned
  near the N terminus so they have more time to fold during translation.
  A phenomenological model predicts that CTF can be accelerated by asymmetry, up to double the rate,
  when folding time is commensurate with translation time; analysis of the PDB
  reveals that structural asymmetry is indeed maximal in this regime.
  This correspondence is greater in prokaryotes, which generally
  require faster protein production.
  Altogether, this indicates that accelerating CTF is a substantial evolutionary
  force whose interplay with stability and functionality is encoded in sequence asymmetry.
}

\section*{Introduction}
  All proteins are translated sequentially from the N- to the C-terminus,
  and are thus inherently asymmetric \cite{saljb65}. 
  One example of such N-to-C asymmetry is signal peptides, which enable
  translocation across membranes, and are located at the N terminus \cite{vonjm90}. 
  This raises the general question of whether and how
  asymmetry in protein production is leveraged to gain evolutionary advantage.
  Here we examine structural data from the protein data bank (PDB)
  in search of traces of such adaptation.
  We analyzed the distribution of secondary structure along the sequence for \num{16200}
  PDB proteins, finding two striking patterns of asymmetry. First, disordered residues are
  principally located at the ends of sequences, and depleted towards the middle.
  Second, $\beta$-strands are enriched
  by \SI{55}{\percent} near the N terminus, while $\alpha$-helices are enriched
  by \SI{22}{\percent} at the C terminus. 
  These findings agree qualitatively with previous reports
 \cite{lobpc10,lobmo20,thona82,bhaac02,marprl05,kripn05,laips06,saubj11,baisr19}.
  This $\ab$~asymmetry peaks at intermediate
  values of sequence length and contact order -- which both correlate negatively with folding rate --
  indicating a possible link between secondary structure asymmetry and folding.
  
  Hence, we further explore the possibility that 
  $\ab$~asymmetry may accelerate protein production, and is therefore a signature of evolutionary adaptation.
    Production of functional proteins from mRNA comprises two concerted processes: 
  directional translation and cooperative folding. 
  The rate of translation is limited
  by trade-offs between speed, accuracy and dissipation \cite{hoppn74,ninbi75,drunr09,pinpr20}.
  Folding quickly has certain advantages: 
  unfolded proteins lead to aggregation, putting a
  significant burden on the cell \cite{dobtb99,lopce13,sanpn19};
  faster folding allows quicker responses to environmental changes
  \cite{sprmc10,dennr11}. 
  Moreover, organisms whose fitness depends on fast self-reproduction would
  benefit from accelerated protein production that can shorten division time \cite{dilpn11,zubnc14}.
  Proteins may begin folding during translation, although the extent to which this
  varies across the proteome is unclear
  \cite{aleps93,evajm08,holsc15,kimsc15,wrupn17,nilns17,samsa18,liubi20,zhaco11,troar16,cabns16,wautb19}, 
  and depends on the collective dynamics of the folding process.
  Thus, in principle, faster production times
  may be achieved if proteins finish folding and translation at around the same time. 
  This co-translational folding (CTF) enables adaptations that increase
  yield and kinetics of protein production \cite{liubi20,zhaco11,troar16,kraab19,wautb19}.
  For example, nascent peptides interact with ribosomes and chaperones to reduce aggregation
  and misfolding \cite{kaina06,kaisc11,obrja12,obrnc14,liumc19}, 
  while translation rates can be tuned to facilitate
  correct folding \cite{kimsc15,jacpn17,bitpn20,walpn20}. 
  Specifically, we ask if structural asymmetry may have evolved for fast and efficient production via CTF.

  \begin{figure*}[t!]  \centering
  \includegraphics[width=0.99\linewidth]{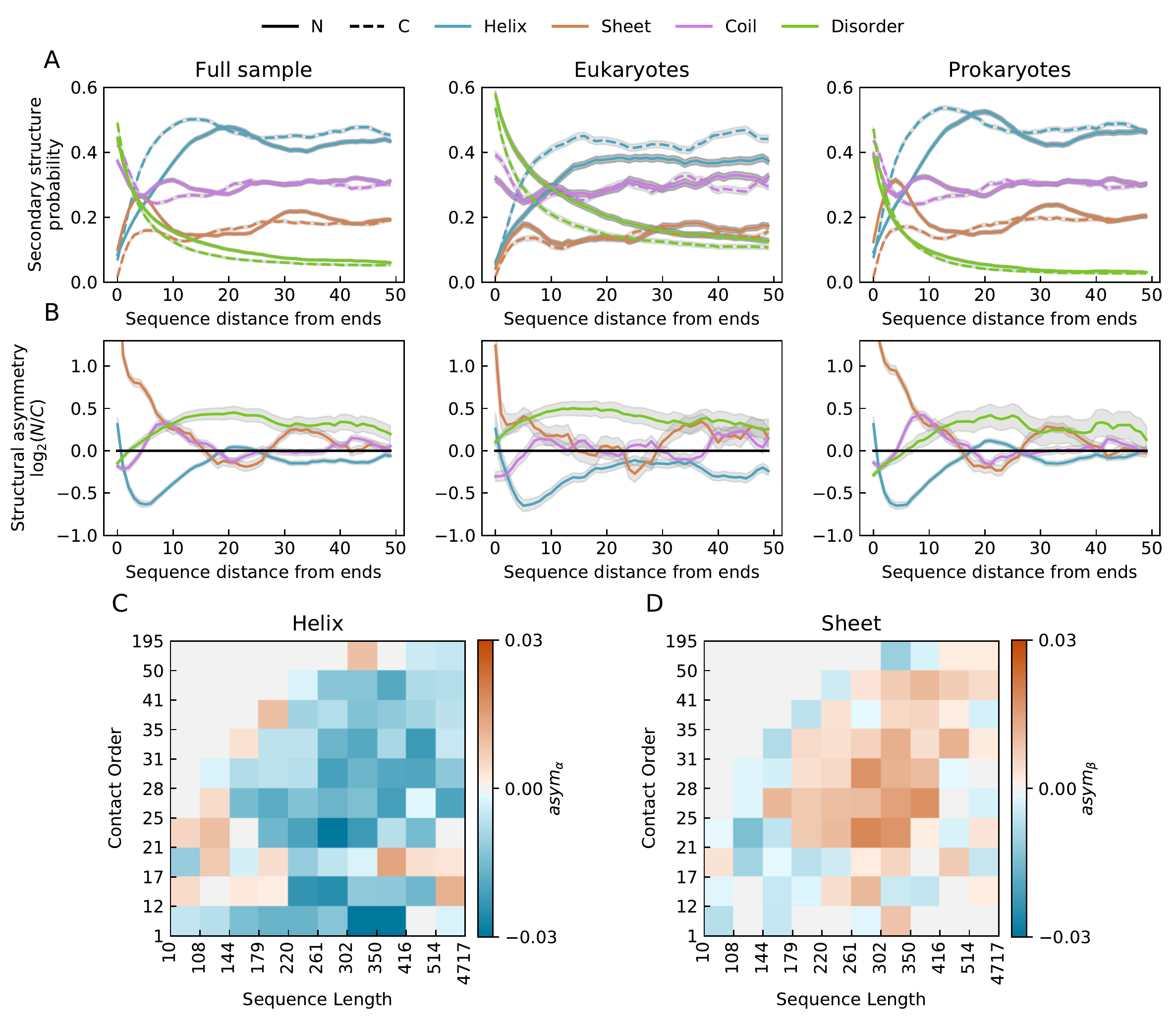}
  \caption{\label{fig:fig1}
  (A) Distribution of secondary structure along the sequence as a function of distance from the N- and
  C-terminus, and (B) the structural asymmetry -- the ratio of the N and C distributions (in
  $\log_2$ scale; $\pm1$ are 2:1 and 1:2 N/C ratios) -- for all \num{16200} proteins (left),
  \num{4702} eukaryotic proteins (middle), and \num{10966} prokaryotic proteins (right). 
  Shading indicates bootstrapped \num{95}$\%$ confidence intervals in both A and B.
  C-D: Mean $\asyma$ (C) and $\asymb$ (D)
   as a function of sequence length and contact order (Eq. \ref{eq:CO}). The data is split
  into deciles and the bin edges are indicated on the axes.
  To reduce noise in the figure due to undersampling, we only colour bins where there
  are 20 or more proteins (full data is shown in SI Fig 1).
  }
  \end{figure*}

  We show that the structural asymmetry observed in proteins is consistent
  with a scheme for accelerating CTF based on the directional
  nature of translation and the heterogeneity of folding rates along the
  sequence \cite{ferco97,ensjm07,linsc11,engpn14,malps16,jacbj16}
  -- \eg~cooperatively-folding protein segments containing $\beta$-sheets
  may fold slower compared to those containing $\alpha$-helices \cite{morpo04}.
    In the proposed \textit{slowest-first} scheme, protein sequences take
  advantage of this heterogeneity by evolving structural asymmetry:
  the slowest-folding segments are enriched at 
  the N-terminus \cite{thona82,bhaac02,marprl05,kripn05,laips06,saubj11,baisr19}, so that they are translated first and have more time to fold. 
  This scheme applies to proteins composed of several independently-folding segments. 
  Therefore, due to the cooperative nature of the many-body folding process, 
  the scheme is more likely to apply to larger proteins.

  A simple model
  predicts that, under the slowest-first scheme, production rate can be almost doubled when
  folding time is equivalent to translation time. 
  To examine this hypothesis, we estimate the ratio of folding to translation
  time of the PDB proteins and compare it with their $\ab$~asymmetry, 
  finding that asymmetry peaks when folding time is commensurate
  with translation time. 
  In this region, proteins are twice as likely to exhibit $\ab$~asymmetry 
  that favours the slowest-first scheme.
  We see more evidence for this scheme in prokaryotic proteins, 
  which is consistent with prokaryotes' greater need for fast protein production due to more frequent cell division.
  Taken together, these findings suggest that proteins sequences have been
  adapted for accelerated CTF via structural asymmetry.

  \begin{figure*}[t!]  \centering
  \includegraphics[width=1.00\linewidth]{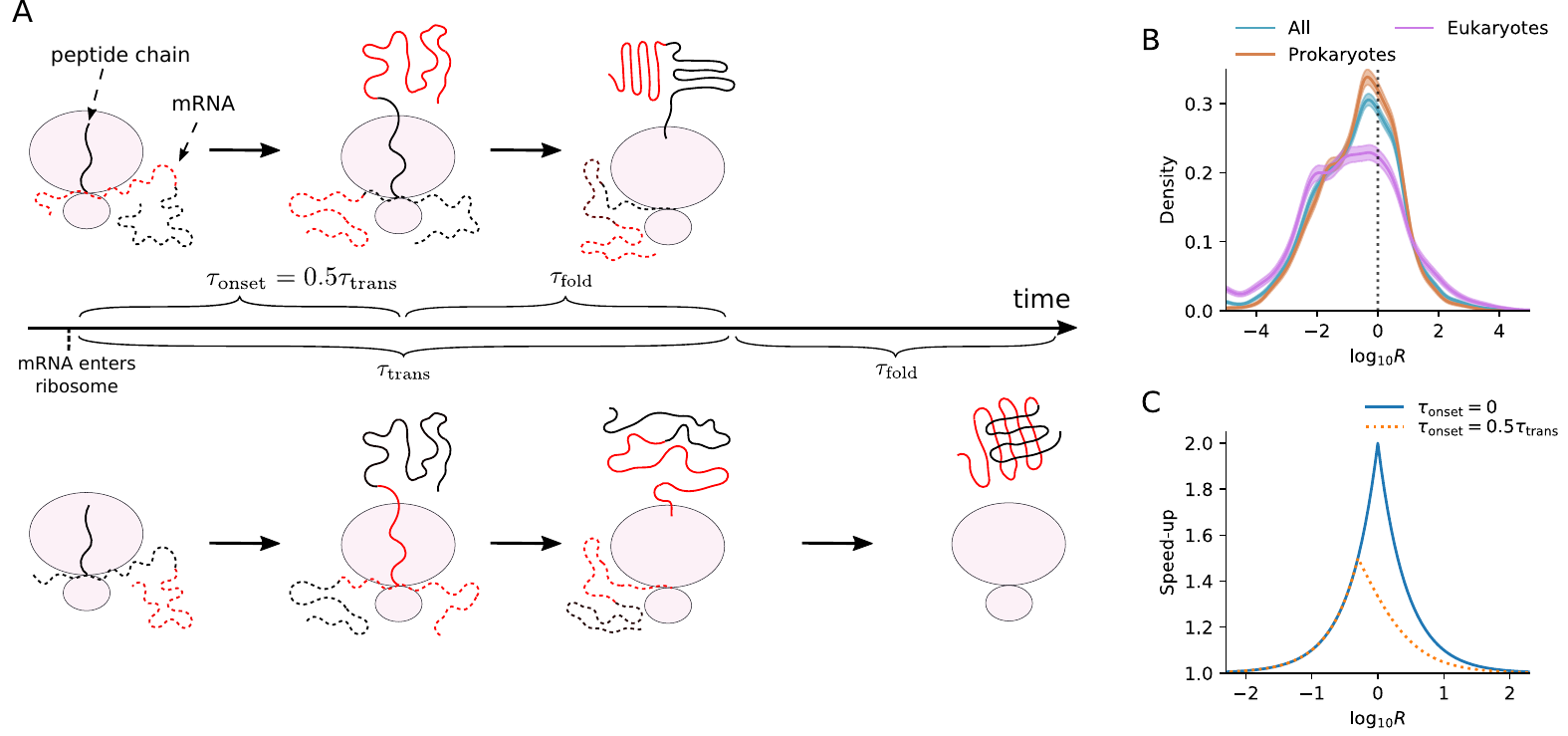}
  \caption{\label{fig:fig2}
  \textbf{Co-translational folding and the slowest-first mechanism.}
  \: A: A section of mRNA (red), encoding a protein segment capable of 
  folding cooperatively, is translated starting from the N terminal side (left).
  The protein segment translocates through the ribosome channel (middle), 
  and undergoes folding once the full segment has been translated and is
  free from steric constraints (right). The time from when the segment is about
  to undergo translation, until the onset of folding, is labelled $\tonset$.
  If this segment is a bottleneck for protein folding, the protein
  will fold faster when this segment is located at the N terminus (top)
  instead of the C terminus (bottom).
  B: Distribution of $R = \tfold / \ttrans$, the ratio of folding to translation
  time (Eq. \ref{eq:R}), for our entire sample, prokaryotic proteins, and eukaryotic proteins.
  Solid lines are kernel density estimation fits to histograms;
  dotted line indicates $R=1$; shading indicates bootstrapped \SI{95}{\percent} confidence intervals.
  C: Theoretical maximum speedup of production rate as a function of $R$ 
  and $\tonset$ (Eq. \ref{eq:speedup}).
  }
  \end{figure*}

\subsection*{Results}

  \subsubsection*{Protein secondary structure is asymmetric}
  Given the vectorial nature of protein translation, 
  one may expect corresponding asymmetries in protein structure. 
  To probe this, we study a non-redundant set of \num{16200} proteins
  from the Protein Data Bank (PDB) \cite{berna00}.
  We find that these PDB proteins exhibit significant asymmetry
  in secondary structure (Fig. \ref{fig:fig1}A-B), even when counting
  each SCOP family once (SI Fig 2). For example, the first $20$ residues at the
  N terminus are on average \SI{55}{\percent} more likely to form strands, and
  the first $20$ residues at the C terminus are \SI{22}{\percent} more
  likely to form helices (in Fig. \ref{fig:fig1}B this is reported in
  $\log_{2}$ scale, but we report $\%$ here). 
  This asymmetry is stronger for prokaryotic proteins
  (\SI{72}{\percent}; \SI{20}{\percent}) than for eukaryotic proteins (\SI{20}{\percent};
  \SI{28}{\percent}). The substantial $\ab$~asymmetry points to an
  evolutionary driving force which we further investigate.
  
  In both N- and C-termini, the $\alpha$-helix and $\beta$-strand distributions
  have a well-defined shape: they
  exhibit \textit{periodicity} in the positioning of these elements along the sequence.
  This periodicity is matched by several $\alpha\beta$-type protein folds where
  $\alpha$-helices and $\beta$-strands are arranged in alternating order
  (SI Fig. 3). These folds tend to be more abundant in prokaryotic proteins (SI Table 1);
  for example, ferredoxin-like folds exhibit high $\ab$~asymmetry, significant
  periodicity at the N-terminus, and are $\sim3$ times more common in prokaryotes.

  The distribution of disordered regions exhibits a different pattern of asymmetry:
  disordered residues are enriched at both ends of proteins compared to the middle
  \cite{lobpc10,lobmo20}.
  Eukaryotic proteins are significantly more disordered, where the probability of
  disorder is well approximated by $\sim D^{-0.5}$, where $D$ is the distance from the
  end, while in prokaryotic proteins the probability of disorder decays as $\sim D^{-1}$.
  Proteins also tend to be more disordered at the N termini \cite{lobpc10}: eukaryotic proteins
  are $\SI{30}{\percent}$ more likely to be disordered within the first \num{100} residues
  of the N terminus compared to the C terminus (prokaryotes: $\SI{17}{\percent}$).
  Although prokaryotic proteins are less disordered than eukaryotic ones,
  the ratio of the numbers of residues in $\beta$ strands and $\alpha$ helices is the same.

  \subsubsection*{Structural asymmetry correlates with sequence length and contact order}
  To better understand the $\ab$~asymmetry, we examined correlations with
  sequence length, $L$, and contact order, CO.
  CO is the average sequence distance between intra-protein contacts,
  \begin{equation}
  \label{eq:CO}
    \textrm{CO} = \big\langle \lvert j - i \rvert \big\rangle~,
  \end{equation}
  where $i$ and $j$ are pairs of residue indices for each contact \cite{ivaps03}.
  High CO is likely to result in greater entropy loss between the unfolded and transition state,
  thus increasing folding time.
  
  To quantify secondary structure asymmetry,
  we calculate the magnitude of asymmetry normalized by length,
  \begin{align}
    \asyma = &\left(N_{\alpha} - C_{\alpha}\right) / L~, \\
    \asymb = & \left(N_{\beta} - C_{\beta}\right) / L~, \nonumber
  \end{align}
  where $N_{\alpha}$ ($N_{\beta}$) and $C_{\alpha}$ ($C_{\beta}$) are the number of residues
  in a $\alpha$-helices ($\beta$-strands) in the N and C halves of a protein sequence.

  \begin{figure*}[t]  \centering
  \includegraphics[width=0.99\linewidth]{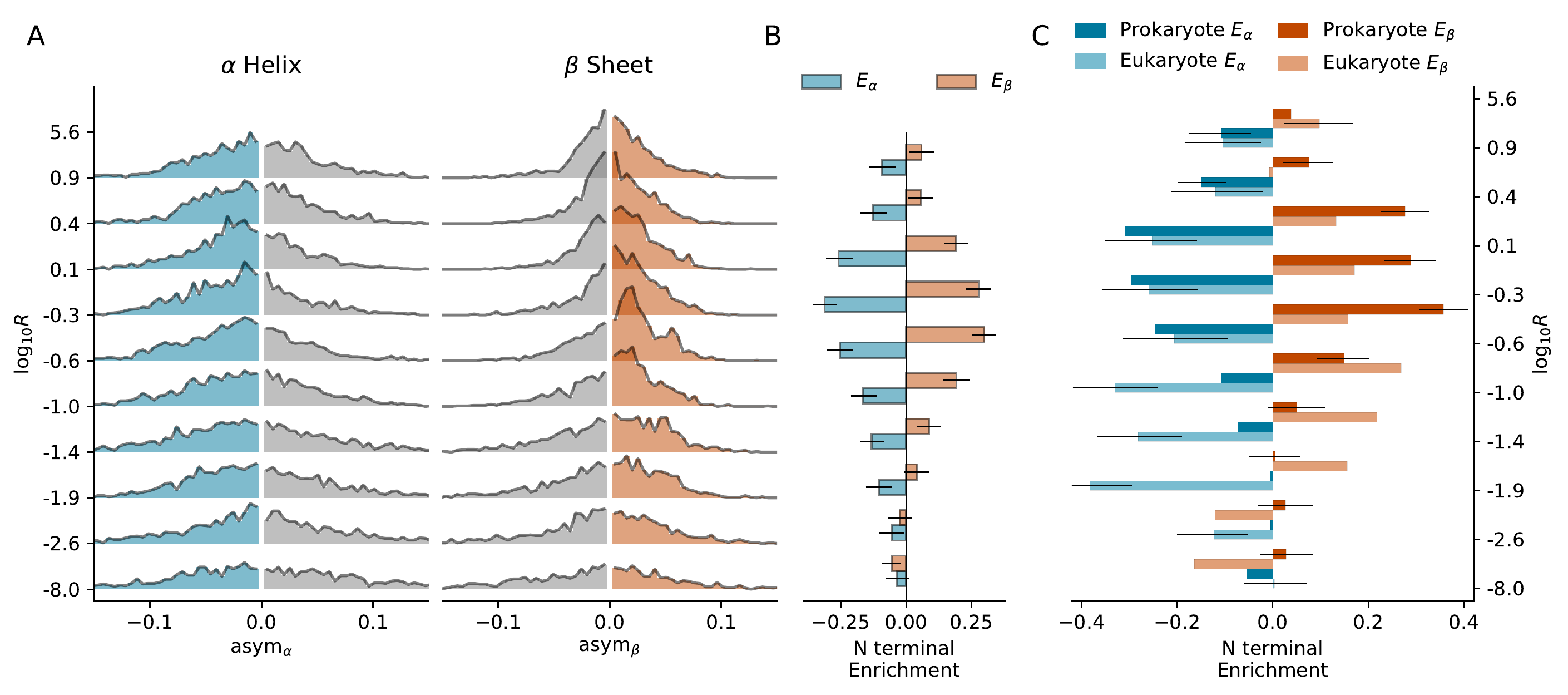}
  \caption{\label{fig:fig3}
  A: $\ab$~asymmetry distributions as a function of $R$, the folding/translation time ratio (Eq. \ref{eq:R}).
  Proteins are divided into deciles according to $R$;
  bin edges are shown on the y-axis.
  B: N terminal enrichment -- the degree to which strands/helices
  are enriched in the N over the C terminus (Eq. \ref{eq:enrich}) --
  is shown for the deciles given in B.
  C: N terminal enrichment as a function of $R$ for \num{4702} eukaryotic proteins
  and \num{10966} prokaryotic proteins. Proteins are divided into bins according to $R$;
  bin edges, shown on the y-axis, are the same as in A-B.
  Whiskers indicate bootstrapped $95\%$ confidence intervals.
 }
  \end{figure*}

  We find that $\ab$~asymmetry is a non-monotonic function 
  of both $L$ and CO (Fig. \ref{fig:fig1}C-D; this is to some
  extent expected since $L$ and CO are correlated, $r=0.65$).
  In particular, there is a region of intermediate length ($179 - 416$)
  and intermediate CO ($21-35$) where structural asymmetry is most apparent (SI Fig. 4).
  The fact that both quantities correlate negatively with
  folding rate ($L$, $r=-0.68$; CO, $r=-0.64$; SI Fig. 5) \cite{nagja05,ivaps03,mansr19,zhabj20},
  taken together with proteins' inherent asymmetry due to
  vectorial translation, leads us to suspect that the origins of this
  $\ab$~asymmetry may be related to co-translational folding.

\subsubsection*{Co-translational folding appears to be widespread}

  During protein production, the ribosome advances along the mRNA from
  the N to the C terminus (Fig. \ref{fig:fig2}A). Proteins may fold in stages during
  translation \cite{evajm08,notjm18,nilns17,wrupn17,kemjm19,holsc15,merbi18,liuel20,elfps20,khumc11,kimsc15}, but the extent to which this happens in still unclear
  \cite{cabns16,waupn18,kemjm19,eicpn10,guipn18,marjm18,kaisc11,tiapn18}.
  In principle, one way to maximise the rate of production and to minimise
  aggregation is by making proteins fold faster than they are translated, or at
  a similar rate. We can obtain a rough approximation of how often this occurs
  by estimating folding rates and translation rates of proteins.
  We estimate the folding rate $\kfold$ using a power law scaling with length (not CO)
  fitted to data from the protein folding kinetics database (PFDB) \cite{mansr19} (Methods). 
  We assume an average translation rate $\ktrans$ that depends on the organism.
  Thus we can estimate the ratio $R$ of folding time $\tfold$ to translation time $\ttrans$,
  \begin{equation}
  \label{eq:R}
    R = \frac{\tfold}{\ttrans}=\frac{1/\kfold}{L/\ktrans} ~.
  \end{equation}
  
  The estimated $R$ distribution exhibits a peak in the region of commensurate times
  $R \approx 1$ (Fig. \ref{fig:fig2}B).  For the \SI{68}{\percent} of proteins (CI
  $\SIrange{53}{88}{\percent}$, SI Fig. 6) that lie in the region $R \le 1$, folding may
  be quicker than translation, suggesting that co-translational folding (CTF) is common.  
  In comparison, a more rigorous method estimated that in \SI{37}{\percent} of proteins in
  \textit{E. coli}, at least one domain will fully fold before translation finishes \cite{cirpn13}.
  Examining prokaryotic proteins and eukaryotic proteins separately reveals a sharper peak
  in the $R$ distribution for prokaryotic
  proteins in the region of commensurate folding and translation times, $1/10<R<10$.
  Notably, a greater fraction of prokaryotic proteins (\SI{56}{\percent}) are in this
  regime compared to eukaryotic proteins (\SI{41}{\percent}).

\subsubsection*{Folding rate asymmetry can speed up co-translational folding}

  Fig. \ref{fig:fig2}A shows two scenarios where the folding rate $\tfold$ is
  determined by a rate-limiting fold (red segment) \cite{frijm03,jacbj16,hansr16,tiapn18}. 
  This bottleneck represents an independent, cooperatively-folding
  segment (\ie~ a `foldon' \cite{engpn14}) that folds slowly.
  If the bottleneck is located at the N terminus (top in Fig. \ref{fig:fig2}A),
  then the production time is minimal, $\tmin = \max (\tfold + \tonset,\ttrans),$
  where $\tonset$ is the time it takes for the segment to be translated, pass through
  the ${\sim}10$ nm~long ribosome tunnel and be free of steric constraints
  \cite{liubi20,zhabj20,notjm18,liupn19}.
   In the other extreme, if the rate-limiting fold includes the C terminus
   (bottom in Fig. \ref{fig:fig2}A), production time is maximized \cite{chenc20},
$  \tmax = \tfold + \ttrans.$
  In this case, the last element can escape the ribosome quickly after being
  translated since it is not delayed by downstream
  translation \cite{nisja20}. Thus, production rate can be accelerated by a factor,
  \begin{equation}
  \label{eq:speedup}
    \textrm{speedup} = 
    \frac{\tmax}{\tmin}~. 
  \end{equation}
  In the limit $\tonset \ll \ttrans$, one finds from Eqs. \ref{eq:R}-\ref{eq:speedup}
  that the speedup as a function of $R = \tfold/\ttrans$ (Fig. \ref{fig:fig2}D) is
  \begin{equation}
  \label{eq:speedupR}
  \textrm{speedup} = 1 + e^{-\lvert \ln{R} \rvert}~.    
  \end{equation}
  A maximal, twofold speedup is achieved when translation time equals folding time, $R=1$,
  and taking $\tonset>0$ shifts this maximum towards $R<1$. 
  
  In practice, $\tonset$ depends
  on the size of the cooperatively-folding protein segment. 
  It has been suggested that
  such a `foldon' can have as few as 20 residues \cite{engpn14,hupn16}, 
  but in general, cooperatively-folding units appear to be larger than this
  -- proteins that are observed to fold via two-state kinetics  can vary greatly in size \cite{depjm95,fer99,jonja03}. 
  Thus, this speedup may only be applicable to larger proteins that fold via multi-state kinetics; this is compatible
  with the prediction that a speedup is substantial only when $\ttrans \approx \tfold$.

  \subsubsection*{Structural asymmetry is maximum for commensurate folding and translation times}
  The speedup curve (Fig. \ref{fig:fig2}D) implies that proteins can benefit
  the most from structural asymmetry when $R = \tfold/ \ttrans \approx 1$.
  Hence, we estimate the magnitude of $\ab$~asymmetry
  as a function of $R$, and plot the distributions in Fig. \ref{fig:fig3}A.
  We note that $R$ is not estimated very precisely, and thus it is useful
  to average out the errors by examining the proteins in large bins (deciles).
  At intermediate $R$, the means of the distributions shift away from
  zero, indicative of strong bias.

  To capture the magnitude of these shifts we calculate the N terminal enrichment, $E$, defined 
  as the fraction of proteins with positive asymmetry (\ie~enriched at the N terminus) 
  minus the fraction of proteins with negative asymmetry (enriched at the C terminus),
  for both helices and strands:
 \begin{align}
 \label{eq:enrich}
    E_{\alpha} = & P(\asyma>0) - P(\asyma<0), \\
    E_{\beta} = & P(\asymb>0) - P(\asymb<0).   \nonumber
 \end{align}

  Fig. \ref{fig:fig3}B shows that in the $R$ deciles with maximum asymmetry,
  proteins in the PDB are $2.0$ times as likely to
  be enriched in $\beta$-strands in the N terminus,
  while $\alpha$-helices are $1.9$ times more likely to be enriched in the C-terminal half. 
  This maximum is found when $-0.6 \leq \log_{10} R \leq 0.1$ (the \SI{95}{\percent}
  confidence intervals for the $\kfold$ estimate give $-1.1 \leq \log_{10} R \leq 0.7$;
  SI Fig. 7).
  This region of maximal asymmetry overlaps with the region of maximal
  speedup (Fig. \ref{fig:fig2}D, Eq. \ref{eq:speedupR}), suggesting that
  asymmetry evolves because it enhances CTF.
  To help put these figures in context, in the pentile where proteins 
  have maximum asymmetry ($-0.6 \leq \log_{10}R \leq 0.1$),
  $11\%$ of proteins have $\textrm{asym}_{\beta}\geq0.05$ (which,
 for the mean length, $L=303$, and an average $\beta$-strand length of $5$ residues, 
  corresponds to about three extra $\beta$-strands near the N terminus)
  while half as many have $\textrm{asym}_{\beta}\leq-0.05$ (the opposite case) (SI Fig 8).

  \subsubsection*{Prokaryotes exhibit greater asymmetry than Eukaryotes}
  We looked at $\ab$~asymmetry for prokaryotic and eukaryotic
  proteins separately, finding that when asymmetry is maximum,
  prokaryotes exhibit more asymmetry than eukaryotes -- strands
  are \SI{36}{\percent} more likely to be enriched at
  the N terminus in prokaryotes compared to eukaryotes (Fig. \ref{fig:fig3}C).
  Typically, prokaryotic cells divide more frequently than eukaryotic cells \cite{zubnc14}, 
  and thus have a greater need for fast production of functional proteins. 
  The analysis is therefore consistent with the slowest-first scheme
  that implies that the stronger pressure on prokaryotes should lead to
  greater asymmetry.

  \begin{figure}[h!]  \centering
  \includegraphics[width=1.00\linewidth]{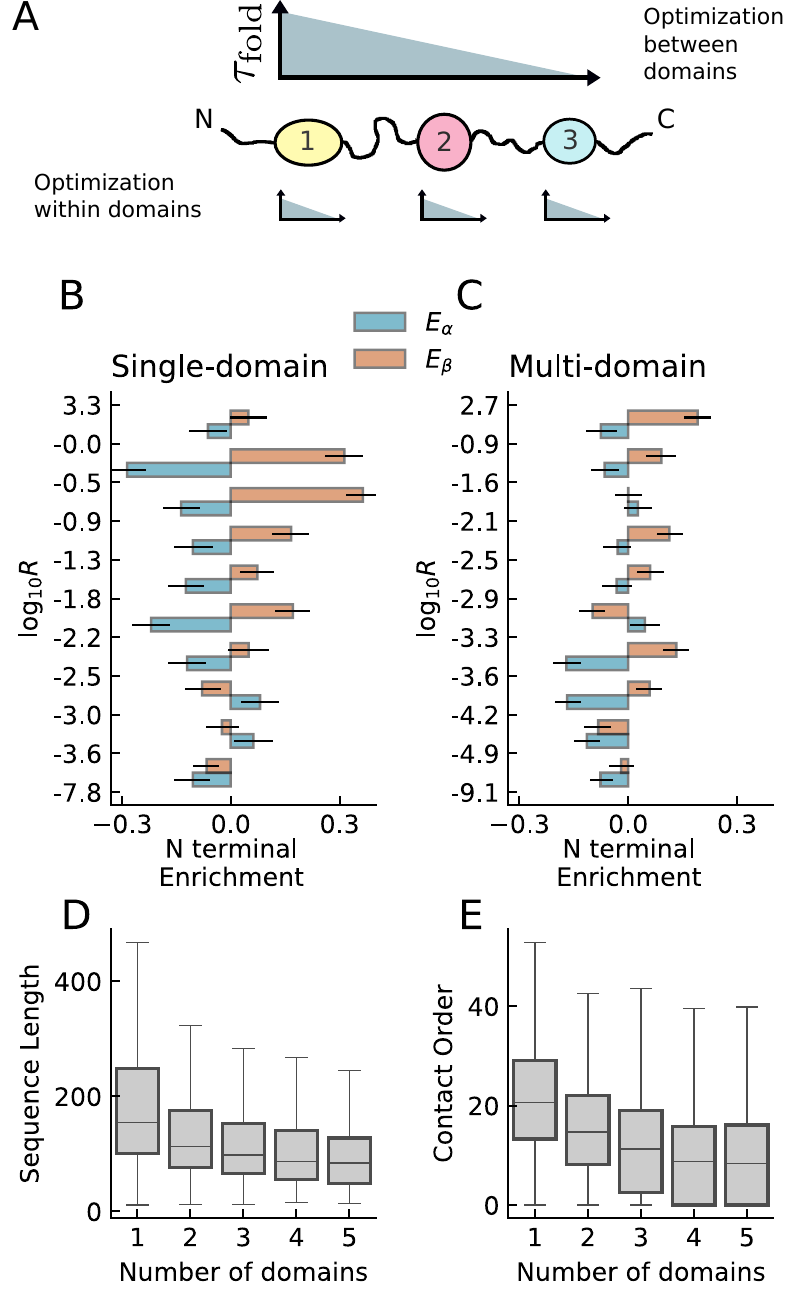}
  \caption{\label{fig:fig4}
  A: Multi-domain proteins can be optimized via asymmetry
  between domains, and/or within domains.
  B-C: N terminal enrichment within domains as a function of $R$
  for single-domain proteins (B: \num{14442} domains) and multi-domain proteins
  (C: \num{23832} domains). Domains are split into deciles based on $R$, and
  the bin edges are shown on the y-axis; whiskers indicate bootstrapped
  $95\%$ confidence intervals.
  D-E: Domain sequence length and contact order distributions for proteins with different
  numbers of domains.
  }
  \end{figure}

  \subsubsection*{Multi-domain proteins are optimized for CTF via distinct mechanisms}
  Multi-domain proteins can be potentially adapted at two levels: within domains,
  and between domains (Fig. \ref{fig:fig4}A).
  To test this, we isolated individual domains in the PDB (using Pfam) \cite{elgna19},
  and calculated CO and $\ab$~asymmetry for each domain as in Fig. \ref{fig:fig3}.
  While intra-domain optimization of secondary structure clearly occurs within single-domain proteins, it is much weaker within multi-domain proteins (Fig. \ref{fig:fig4}B-C).
  Inter-domain optimization entails ordering the slowest-folding domains at
  the N terminus, for which we find no significant bias (SI Fig. 9).
  Instead, we find that as the number of domains increases,
  the size and CO of individual domains decreases (Fig. \ref{fig:fig4}D-E). 
  Thus CTF is maintained in multi-domain proteins mostly 
  by using faster-folding domains throughout.

\subsection*{Discussion}

  We examined the hypothesis that proteins are selected for CTF to hasten protein production and reduce aggregation/misfolding, but this may not be equally true for all proteins.
  As an example, we showed that in prokaryotes, which have a greater burden of
  cell growth, proteins tend to have more asymmetry than in eukaryotes.
  Topological constraints may preclude early folding of the N terminus
  if the C-terminus is also part of the folding nucleus \cite{tiapn18}.
  More generally, CTF may be hindered in some proteins by interactions with the
  ribosome \cite{kemjm19}. Long-lived proteins \cite{toynr13} may
  derive little benefit from an increase in production speed.
  On the other hand, proteins produced in large quantities need to fold quickly
  as aggregation can increase non-linearly with concentration \cite{wanij05}.
  Some of these predictions can be tested when sufficient data for protein lifespan
  \cite{fornc18}, expression levels \cite{yunbi07}, and structure become available. 
  While we showed that $\ab$ asymmetry is apparent in a broad set of 
  proteins, further analysis of an extended data set may be able to detect the sub-classes
  of proteins that will benefit most from $\ab$ asymmetry.

  Our model proposes an absolute maximum of a twofold acceleration
  in protein production, which at first glance, may seem insignificant compared
  to the range of protein folding times ($\sim 9$ orders
  of magnitude). However, we show that this acceleration is only
  possible when folding time is similar to translation time (\ie~ on the order of seconds to minutes).
  We expect that, even for a single protein,  decreasing production time
  by seconds to minutes can increase fitness, and when this
  is applied over a large group of proteins this should be quite substantial.

\begin{figure}[ht]  \centering
  \includegraphics[width=0.99\linewidth]{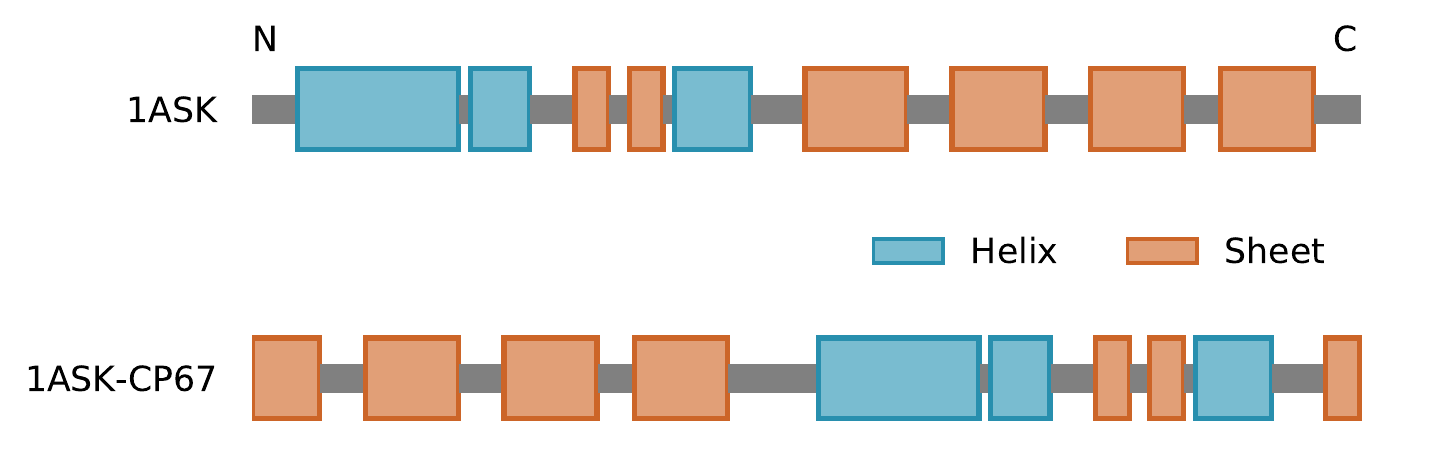}
  \caption{\label{fig:fig5}
  Secondary structure for nuclear transport factor 2 H66A mutant (PDB: 1ASK \cite{clajm97})
  and a circular permutant, 1ASK-CP67, which may fold faster during translation.
  }
  \end{figure}

  To experimentally test the slowest-first mechanism, we suggest studying CTF of multiple proteins with $R \approx 1$, which differ
  in $\asyma$ and $\asymb$. In particular, we propose to use proteins whose 
  sequences are related by \textit{circular permutation}, while having identical structures \cite{milpn02,lona08,kemjm15,marjm18}. 
  Circular permutants with opposite structural asymmetry, as the example in
  Fig. \ref{fig:fig5}, should fold at significantly different rates.  
  Additional experimental control of $R$ is possible via 
  synonymous codon mutations \cite{komfe99} or \textit{in vitro} expression
  systems \cite{samsa18}. Thus, one can test whether asymmetry in secondary structure
  can lead to acceleration of CTF, and how this depends on $R$.

  \subsubsection*{CTF for multi-domain proteins is more complex}
  Multi-domain proteins exhibit less asymmetry than single-domain proteins.
  Due to interactions between domains  \cite{hannr07,notjm18,liumc19,liupn19,bitpn20},
  optimization via asymmetry may not be feasible --- instead, a safe strategy is
  to fold each domain before translating subsequent domains.
  To explain the lack of intra-domain $\ab$ asymmetry (Fig. \ref{fig:fig4}C),
  we propose a simple mechanical argument. When a $\beta$-sheet forms,
  the protein chain contracts. This results in a pulling force on
  both the ribosome \cite{golsc15,leipn19}, and on any upstream domains.
  This extra resistance to $\beta$-sheet formation may preclude the
  early formation of $\beta$-strands at the N terminal side of a domain.
  If this is true, then the domain in position 1 should
  still exhibit $\ab$~asymmetry; we currently lack sufficient statistical
  power to conclusively test this (SI Fig. 10).
  Further tests could look at CTF of a $\beta$-rich domain in the
  the presence or absence of an upstream domain \cite{batjm05,kembi19}.

  \subsubsection*{Disorder is enriched at both sequence ends}
  The N and C termini principally share a notable tendency
  for disorder near the end, which suggests that they are affected by the same physical
  \textit{end effect}. The amino acid at the end is linked to the chain
  by only one peptide bond, leaving it more configurational freedom than
  an amino acid in the centre of the protein, which is constrained by two
  bonds. This entropic contribution to the free energy of the loose ends,
  of order $k_B T$, can induce disorder in marginally stable structures. 
  
  Since disordered regions do not need time to fold, placing them towards the
  C-terminus gives the other residues more time to fold. 
  Yet, we find a similar, slightly stronger, tendency for disorder near
  the N-terminus (green curves in Fig. \ref{fig:fig1}B), particularly in eukaryotes.
  This may result from other determinants of protein evolution;
  \eg, disordered regions tend to interact 
  with some ribosome-associating chaperones \cite{alapb11,wilce13}.
  If disorder at the N terminus is related to chaperones, we expect that
  asymmetry will be higher for slow-folding proteins as they are more prone
  to aggregation. We find that bias for disorder at the N-terminus is strongest for
  slow-folding proteins (high $R$, $L$ and CO; SI Fig.11), but only for
  prokaryotes, not eukaryotes.
  Given the absence of a correlation between $R$, $L$ and CO and disorder asymmetry 
  in eukaryotic proteins, the question of why eukaryotic proteins are
  more disordered at the N terminus remains open.

  \subsubsection*{Considering tertiary structure}
  We used secondary structure as a proxy for folding rate,
  but there are also contributions from tertiary structure.
  To evaluate this assumption, we ran coarse-grained simulations of
  CTF of three structurally-asymmetric proteins
  while varying $R$, for both the original sequence and of the reverse sequence.
  We find that these proteins fold faster when $\beta$-strands
  are translated first, in the relevant region of $R\sim1$ (SI Fig. 12).
  
  We also studied the effect of tertiary structure by looking at asymmetry
  in surface accessibility. $\beta$ strands at the N terminus are less likely
  to be exposed to solvent than $\beta$ strands at the C terminus; 
  this bias is stronger
  for prokaryotic proteins (first 20 residues: $\SI{41}{\percent}$)
  compared to eukaryotic proteins ($\SI{13}{\percent}$) (SI Fig 12).
  Since solvent-exposed $\beta$-strands are less likely to form part of
  a folding nucleus \cite{nolps08}, this suggests that $\beta$-strands
  at the N terminus are more likely to nucleate folding compared to
  those at the C terminus.

  \subsubsection*{Alternatives explanations for asymmetry}
  
  We can think of only few examples of inherent asymmetry in proteins.
  The ends have a slight chemical difference (carbon vs nitrogen), but one atom
  seems insignificant. Amino acids are chiral, but we cannot say how this could lead to
  asymmetry in secondary structure. The only remaining driving force for
  asymmetry, in our view, is the unidrectional translation process.

  One theory exists that predicts asymmetry in $\alpha$-helices:
  Marenduzzo et al. show that in a model of a growing, self-interacting string,
  the string forms helices at the growing (C) end \cite{marprl05}, and suggest that
  this bias affected protein evolution. However, this fails to address
  why there are more beta sheets at the N terminal, nor
  does it explain why asymmetry depends on protein length.

  We could also consider that $\beta$-strands are typically less open to
  solvent, and hence we also see asymmetry in solvent accessibility (SI Fig 12).
  Thus, it is possible that $\beta$-strand asymmetry is a by-product of
  a driving force for having the N terminal side less exposed to solvent.
  The only explanation we can think of for such a driving force is to
  have the folding nucleus located at the N terminus, which is effectively
  covered by the `slowest-first' hypothesis since a folding nucleus can
  be considered a rate-limiting fold.

  \subsubsection*{We need more data on folding rates}
  The data used to fit Eq. \ref{eq:pred} are sparse (\num{122} proteins),
  biased towards small, single-domain proteins, and typically obtained
  from \textit{in vitro} refolding experiments \cite{mansr19}. 
  To test whether  our conclusions are robust to sampling,
  we estimate confidence intervals
  using bootstrapping with sample sizes equal to the original sample size, and
  half that amount; we perform this test on both the reduced version of the PFDB
  data set used in the main figures, and on a second version of the PFDB data
  set (Methods; SI Fig. 7). In addition, we calculate the main results using
  using a different protein folding data set, ACPro \cite{wagps14}, which
  partially overlaps with PFDB, but includes larger proteins (SI Fig. 14).
  In all of the above analyses, the point of maximum asymmetry is found to
  be $1/100 < R < 100$, which corresponds to the region where CTF speed-up is possible.
  However, to fully overcome the aforementioned limitations,
  further experiments are needed.

\subsubsection*{Analysis is consistent with hypothesis that proteins are selected for CTF via secondary structure}
  To sum, in the proposed the \textit{slowest-first} mechanism, CTF can
  be accelerated by positioning the slowest-folding parts of a protein
  near the N terminus so that they have more time to fold. 
  A survey of the PDB shows 
  that the estimated acceleration correlates with asymmetry in secondary structure.
  In particular, the rate of production can be almost doubled when translation
  time is similar to folding time, and indeed these proteins exhibit the maximal asymmetry in secondary structure distribution.
  Altogether, there appears to be substantial evolutionary selection, manifested in sequence asymmetry, for proteins that
  can fold during translation.

\section*{Methods}

\subsubsection*{Data}
  We extracted a set of \num{16200} proteins from the Protein Data Bank (PDB) \cite{berna00}.
  We include proteins where the SEQRES records exactly match the corresponding
  Uniprot sequence (not mutated, spliced, or truncated) \cite{unina18}.
  We include proteins that have been fused or have purification tags, but for
  these we again only include the part that exactly matches the Uniprot sequence.
  For each unique protein sequence, we only include the most recent structure.
  We used SIFTS to map PDB and Uniprot entries \cite{danna18}.
  We exclude proteins with predicted signal peptides as little is known about
  whether such proteins undergo CTF; we used Signal-P5.0 to identify signal
  peptides \cite{alanb19}. Using the above criteria we extracted a set of
  \num{38274} domains by matching PDB entries to Pfam domains \cite{elgna19}.
  We assume that domains identified in Pfam via homology are independent folding units.
  $\alpha$ helix and $\beta$-strands are identified through annotations in
  the PDB; disorder is inferred from residues with missing coordinates.
  To calculate contact order, we only consider contacts between residues
  where $\alpha$ carbons are within  \SI{10}{\angstrom}; we confirm that the correlation in
  Fig. \ref{fig:fig1}D is robust to choice of this cutoff (SI Fig. 15).
  
  We use the protein folding kinetics database (PFDB) for estimating folding
  rates \cite{mansr19}. For our main results we only used entries with 
  realistic physical conditions ( $5 < \textrm{pH} < 8$, and
  $\SI{20}{\celsius}< T < \SI{40}{\celsius}$) and ignored folding rates which
  had been extrapolated to $T = \SI{25}{\celsius}$; in total, \num{122} proteins.
  We test a second version of the PFDB data set without excluding proteins, and using
  folding rates which were extrapolated to $T = \SI{25}{\celsius}$; \num{141} proteins.
  We also use the ACPro data set \cite{wagps14} to test the robustness of our conclusions;
  \num{125} proteins.

\subsubsection*{Predicting folding and translation rate}

 The folding rate, $\kfold$ (in units of 1/sec), is estimated by a
 power-law fit as a function of the protein's length:
  \begin{equation}\label{eq:pred}
    \log_{10} \kfold = A + B\,\log_{10} L ~,
  \end{equation}
  where $L$ is sequence length in residues; $A$ and $B$ are free parameters.
  We fit these parameters using data from the PFDB \cite{mansr19}
  to get \SI{95}{\percent} confidence intervals of $A=13.8 \pm 2.1$
  and $B=-6.1 \pm 1.2$ (with correlation coefficient $r=-0.68$,  and p-value $p<0.005$).
  We consider that proteins that are observed to fold via multi-state kinetics
  may be badly approximated by a single folding rate; we thus repeat the analyses
  with the 89 two-state proteins in the PFDB, finding the area of maximum
  asymmetry within $-1.1 < \log_{10} R < 1.5$.
  The estimate from Eq. \ref{eq:pred} is limited for the following reasons:
  (i) It is extracted from a small set of 122 proteins. (ii) It  disregards
  the effects of secondary structure, contact order order, and other important determinants.
  (iii) The data is from \textit{in vitro} measurements. (vi) The data is biased towards
  small, single-domain proteins.  Thus, it is only a rough predictor for the folding rates of
  \textit{individual} proteins in the set, as the standard deviation between estimated and empirical
  folding rates is $1.22$. For all these reasons, we use Eq. \ref{eq:pred} as an estimator of the \textit{average} folding rate of sets of proteins of similar length $L$ where the large sampling size of each bin is expected to reduce the errors as ${\sim}N^{-1/2}$.

  We tested whether the predicted folding rates of proteins in the PDB are
  within certain approximate bounds on realistic folding rates. 
  A lower bound to folding time has been estimated at ${\sim}L/100$ \SI{}{\us}
  \cite{kubco04}, while we take the doubling time of \textit{E. coli}, roughly
  $20$ minutes, as an approximate upper bound. Of course, many proteins rely on
  chaperones, so their bare estimated folding time may be longer than the upper bound,
  while others come from organisms with much longer doubling times.
  Even so, according to Eq. \ref{eq:pred} only \SI{8}{\percent} of proteins
  are estimated to have a folding time greater than $20$ minutes, while only
  \SI{7}{\percent} of proteins are estimated to fold faster than the lower
  bound. Given the magnitude of the error in estimating the folding time
  of individual proteins, Eq. \ref{eq:pred} appears to yield estimates that
  are mostly within the biologically reasonable regime. 
  Furthermore, in estimating the folding rate of large proteins, a common assumption is that they consist of multiple independently-folding domains \cite{rolja14} -- which
  considerably reduces the estimated folding time of the slowest proteins --
  but we neglect to make this assumption.
  
  In principle, we could have used structural/topological measures (such as contact order, long-range order,
  etc. \cite{sonip10}) to slightly improve the fit to Eq. \ref{eq:pred}. 
  However these typically involve numerous methodological choices and additional parameters
  \cite{wagjt14}, and the scaling  relations are entirely empirical. 
  In contrast, scaling of folding time with length has a robust
  theoretical background \cite{thijp95,gutpr96,ciepr99,kogjm01,lijp02,nagja05,lanjp12,garpn13};
  the exact form of of the scaling is debated, but a power law is favoured slightly \cite{gutpr96,lanpo13}.
  
  We assume the translation rate, $\ktrans$, depends on the organism 
  (host organism for viral proteins), such that $\ktrans$ is $5$ amino acids per second
  for eukaryotes and $10$ for prokaryotes.

\subsection*{Data Availability} 
  The non-redundant sets of proteins and domains, along with the data
  used in the figures and Supplementary Information, will be made available
  on Zenodo.

\subsection*{Code Availability} 
  Simulation and analysis code, along with code used to make all figures, are accessible at
  \url{https://github.com/jomimc/FoldAsymCode}.

\subsection*{Acknowledgements} 
  We acknowledge Albert J. Libchaber for stimulating discussions
  and comments on the manuscript. This work was supported by the
  Institute for Basic Science, Project Code IBS-R020-D1.

\subsection*{Competing Interests} The authors declare that they have no
  competing financial interests.

\subsection*{Correspondence} Correspondence and requests for materials
  should be addressed to J.M.~(email: jmmcbride@protonmail.com)
  and T.T.~(tsvitlusty@gmail.com).

\subsection*{Author Contributions}
  J.M. and T.T. designed research; J.M. performed research;
  J.M. analyzed data; J.M. and T.T. wrote the paper.

\bibliographystyle{unsrtnat}
\bibliography{fold_asym}

\end{document}


\maketitle

\subsection{Coarse-grained simulation of co-translational folding}

  In this work, we assume that $\beta$-sheets fold slower
  than $\alpha$-helices in general, and hence our model predicts that co-translational
  folding (CTF) will finish faster if $\beta$-sheets are located at the N terminal.
  Here we further examine this phenomenon using coarse-grained models of
  proteins, such that we also take into account the effect of tertiary structure.
  We study three proteins that exhibit significant asymmetry in secondary structure,
  with native structures taken from the Protein Data Bank (PDB): 1ILO, 2OT2, 3BID.
  
\begin{figure}[t!]  \centering
  \includegraphics[width=0.95\linewidth]{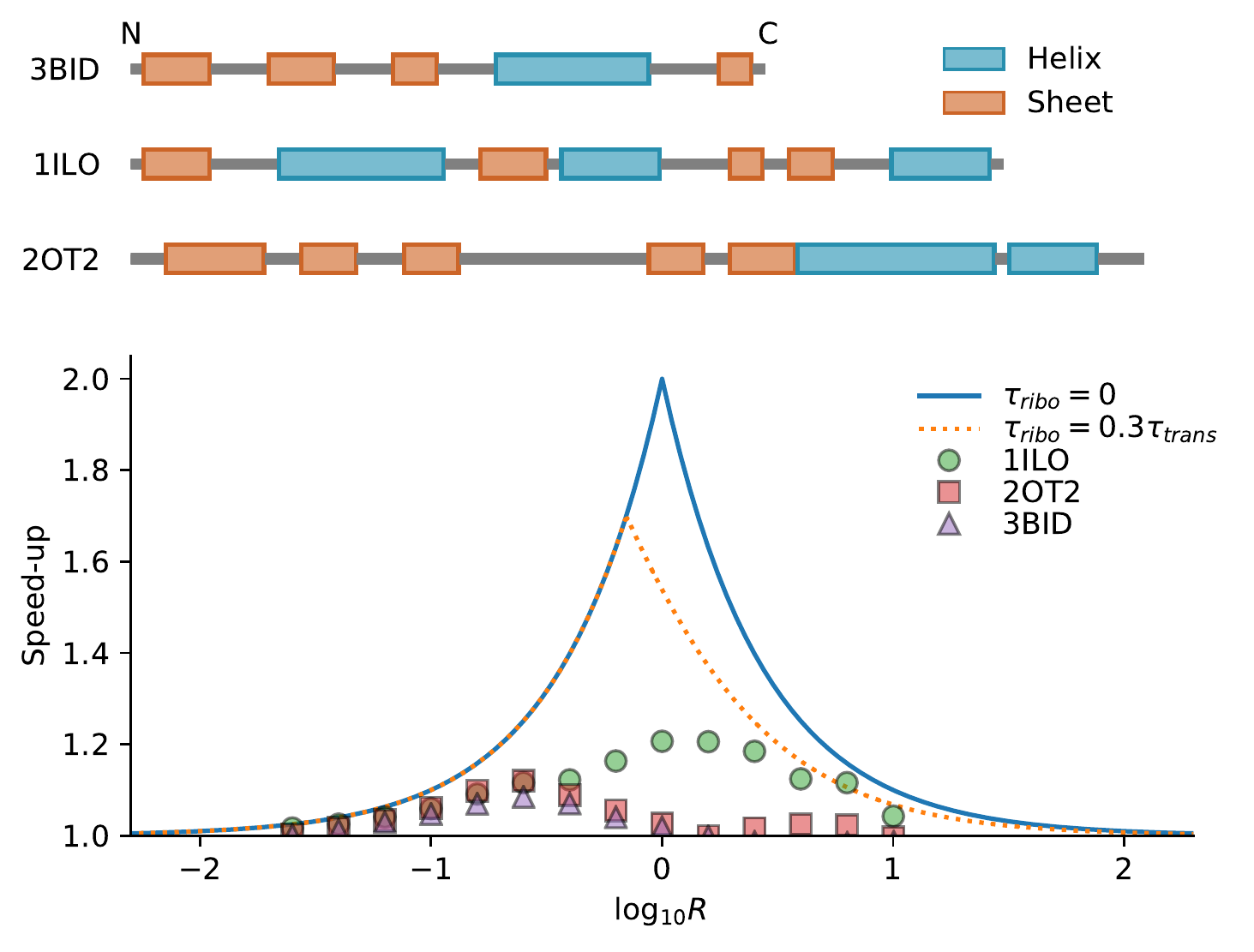}
  \caption{\label{fig:si8}
  (Top) Secondary structure of three proteins that exhibit $\ab$~asymmetry
  in line with the `slowest-first' scheme (PDB IDs: 3BID, 1ILD, 2OT2).
  (Bottom) Maximum theoretical speed-up achievable as a function of $R$
  and $\tribo$ (lines).
  Speed-up achieved in a simulation by translating proteins from the N- to
  the C-terminal (with slow-folding parts near the N terminal)
  compared to translating them from the C- to the N-terminal.
  }
  \end{figure}

  To investigate whether these proteins benefit from having $\beta$-sheets at the N
  terminal, we first calculate the time it takes to fold after starting from
  an unfolded configuration, $\tfold$. Then, for a range of translation times 
  ($\ttrans = L \taa$, where $\taa$ is the time it takes to translate
  one amino acid) we calculate the
  time it takes to undergo translation and folding, $\tctf$, for both translation from the N-
  to the C-terminal ($\tctfn$), and also from the C- to the N-terminal ($\tctfc$).
  We then plot the speed-up, the ratio between the time it takes for CTF
  along the backward to the forward direction ($\tctfc / \tctfn$),
  against the ratio of folding translation time, $R=\tfold / \ttrans$ (Fig. \ref{fig:si8}).
  All three proteins folded faster (up to \num{21}$\%$) when $\beta$-sheets
  were translated first, and the maximum speed-up occurred when $1/10<R<10$.

  We run molecular-dynamics simulations using the Gromacs software package \cite{abrso15}.
  We use a G\={o}-like model forcefield that represents each residue with a single bead,
  generated using the Go-Kit python package \cite{neejc19}.
  To calculate folding time, $\tfold$, we run simulations from the unfolded state
  at $T^*=100$, with a time step of $dt^*=0.001$ for $2\times10^6$ steps; units are
  reported in reduced units according to the definitions given by Gromacs.
  We record folding time, $\tfold$, as the time
  it takes for \num{90}$\%$ of native contacts to be within \num{20}$\%$ of the distance
  given in the PDB structure. We run the simulation \num{1000} times to get an average.
  
  To calculate co-translational folding time, $\tctfn$ ($\tctfc$), we initialise the protein in
  an elongated configuration and restrain all beads except the first bead
  at the N (C) terminal. We run the simulation for $\taa$, the time it takes to
  translate a single residue, and then remove the restraint from the next bead
  in the chain; we repeat until all restraints are removed.
  We also create wall out of purely repulsive spheres packed on a square
  lattice, positioned at the boundary between free and restrained beads,
  which prevents interaction between the `translated' and `untranslated' parts of the
  protein. As each bead has its restraint removed, we move the wall along the chain.
  When the restraint is removed from the final bead, we remove the wall and run
  for a further $5\times10^5$ or $1\times10^6$ time steps. We measure $\tctfn$ ($\tctfc$)
  in the same way as $\tfold$. We run the simulation in each direction \num{4000} times
  to get an average. A full set of simulation parameters is available in the supplementary
  materials in the form of Gromacs input files and python scripts for configuration set-up
  and analysis.
  
  \clearpage

  \begin{table}\centering
  \caption{\label{tab:scop_prob}
  The probability of that a prokaryotic protein has a specific fold, $P_{\textrm{Prok}}$, compared
  with the probability of that a eukaryotic protein has a specific fold, $P_{\textrm{Euk}}$.
  Results are shown for the 9 most common SCOP folds \cite{andna19} identified in PDB data used in this study.
  }
  \begin{tabular}{lllllll}
  SCOP Class & SCOP ID & Count & Description & $P_{\textrm{Prok}}$ & $P_{\textrm{Euk}}$ & Probability ratio  \\ 
  \toprule
  a/b & 2000031 & 171 &  TIM beta/alpha-barrel                     &     0.06  &     0.03  &     1.94 \\
  a/b & 2000148 & 171 &  SDR-type extended Rossmann fold           &     0.06  &     0.04  &     1.33 \\
  a+b & 2000014 & 115 &  Ferredoxin-like                           &     0.04  &     0.01  &     2.85 \\
  a/b & 2000088 & 76  &  Methyltransferase-like                    &     0.02  &     0.02  &     1.10 \\
  a+b & 2000303 & 65  &  Acyl-CoA N-acyltransferases (Nat)         &     0.02  &     0.01  &     4.14 \\
  a   & 2000002 & 62  &  Globin-like                               &     0.00  &     0.07  &     0.07 \\
  a/b & 2000016 & 58  &  Rossmann(2x3)oid (Flavodoxin-like)        &     0.02  &     0.01  &     1.98 \\
  a/b & 2000607 & 58  &  PLP-dependent transferase-like            &     0.02  &     0.02  &     1.04 \\
  a/b & 2000027 & 56  &  ClpP-type beta-alpha superhelix           &     0.02  &     0.00  &      nan \\
  \end{tabular}
  \end{table}

  \clearpage

  \begin{figure}[t!]  \centering
  \includegraphics[width=0.95\linewidth]{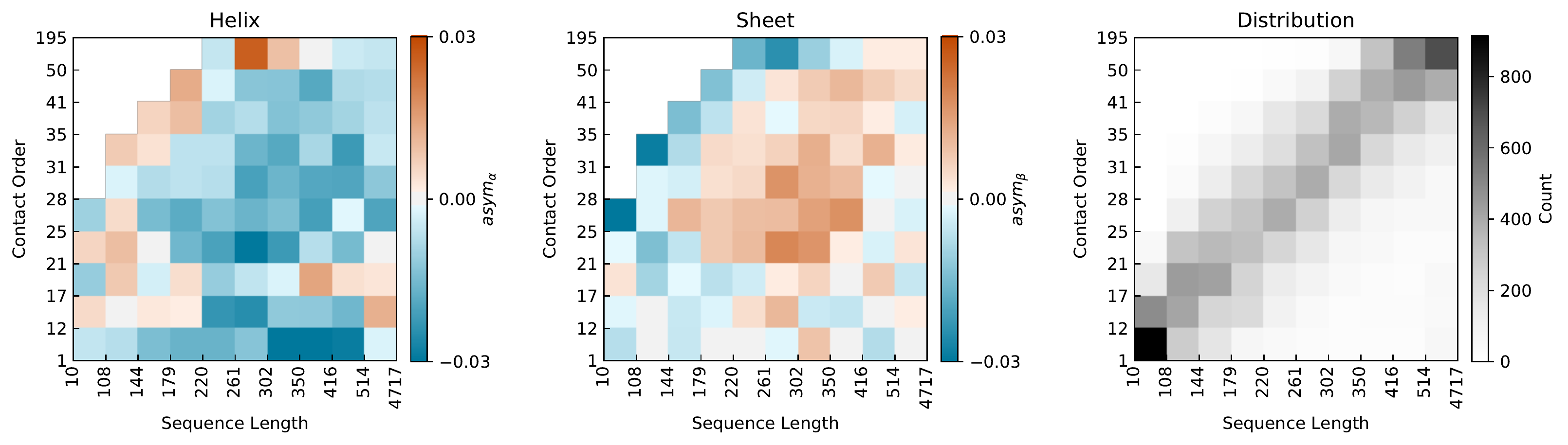}
  \caption{\label{fig:si13}
  A-B: Mean $\asyma$ (A) and $\asymb$ (B)
  as a function of sequence length and contact order. The data is split
  into deciles and the bin edges are indicated on the axes.
  C: Distribution of proteins according to contact order and sequence length.
  }
  \end{figure}

  \clearpage

  \begin{figure}[t!]  \centering
  \includegraphics[width=0.95\linewidth]{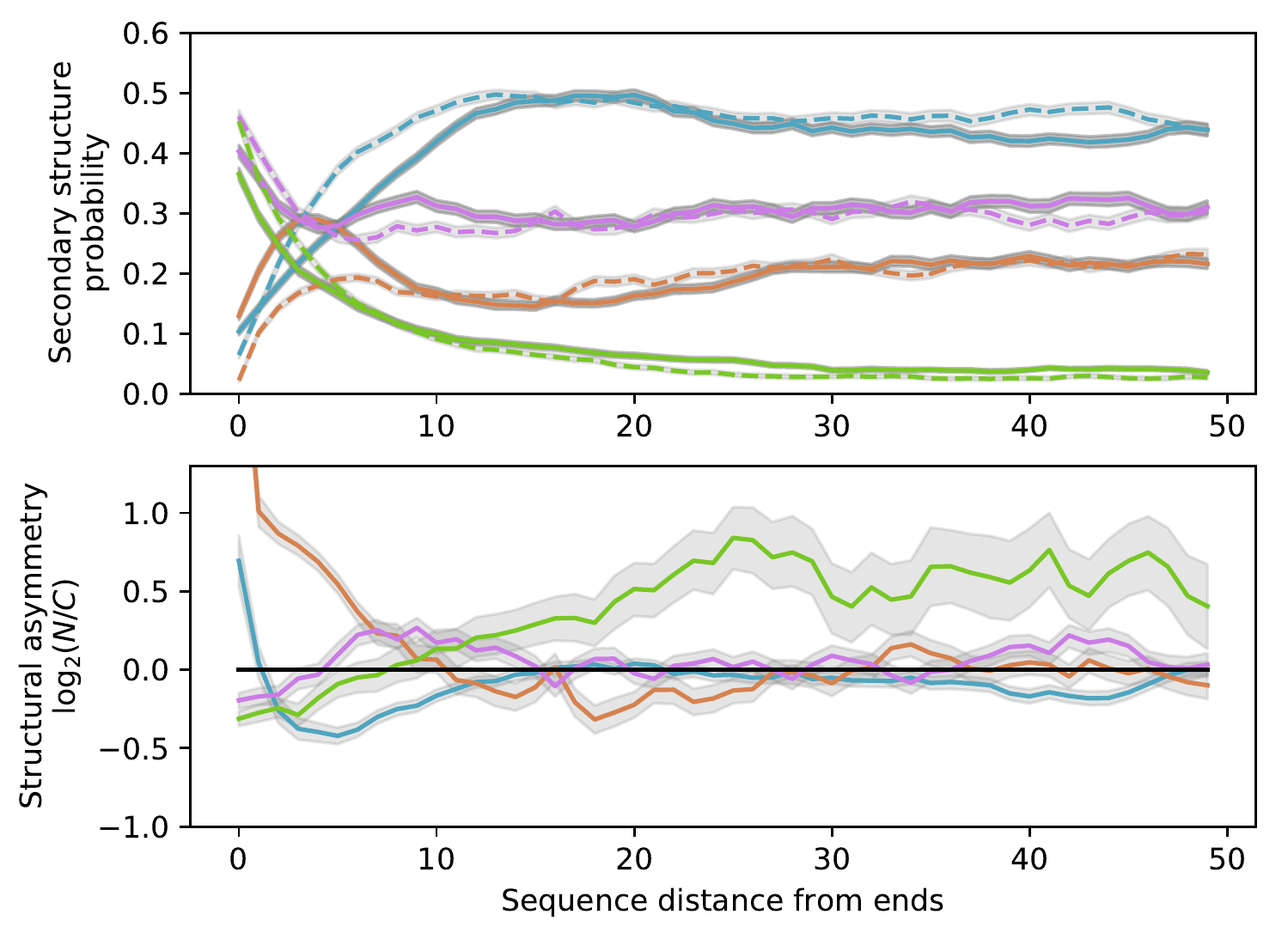}
  \caption{\label{fig:si14}
  (A) Distribution of secondary structure along the sequence as a function
  of distance from the N- and C-terminus, and (B) the structural asymmetry
  -- the ratio of the N and C distributions (in $\log_2$ scale; $\pm1$ are
  2:1 and 1:2 N/C ratios) -- for a reduced set of \num{1526} proteins where
  each SCOP family is included once \cite{andna19}. We were able to match SCOP IDs to
  \num{3290} out of \num{16200} proteins using the Sifts database \cite{danna18}.
  Shading indicates bootstrapped \num{95}$\%$ confidence intervals (A-B).
  }
  \end{figure}

  \clearpage

  \begin{figure}[t!]  \centering
  \includegraphics[width=0.90\linewidth]{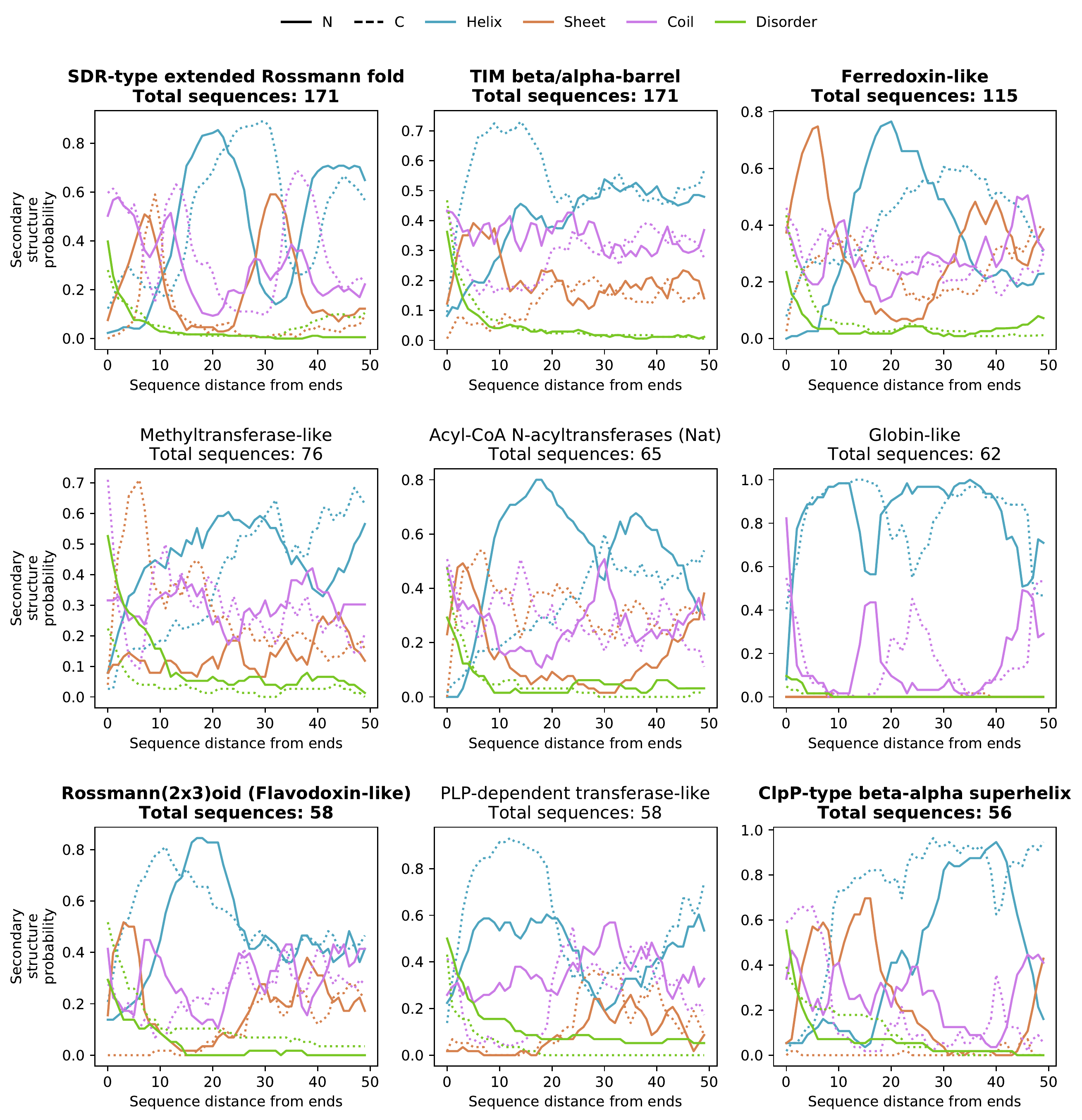}
  \caption{\label{fig:si1}
  Secondary structure probability as a function of sequence distance from the
  N and C terminal respectively, for $\alpha$-helices, $\beta$-sheets,
  random coil, and disorder. Separate plots are shown for the 9 most common
  SCOP folds in our sample of PDB proteins \cite{andna19}.
  The folds that exhibit $\ab$~periodicity in their secondary structure
  are highlighted in bold.
  }
  \end{figure}

  \clearpage

  \begin{figure}[t!]  \centering
  \includegraphics[width=0.95\linewidth]{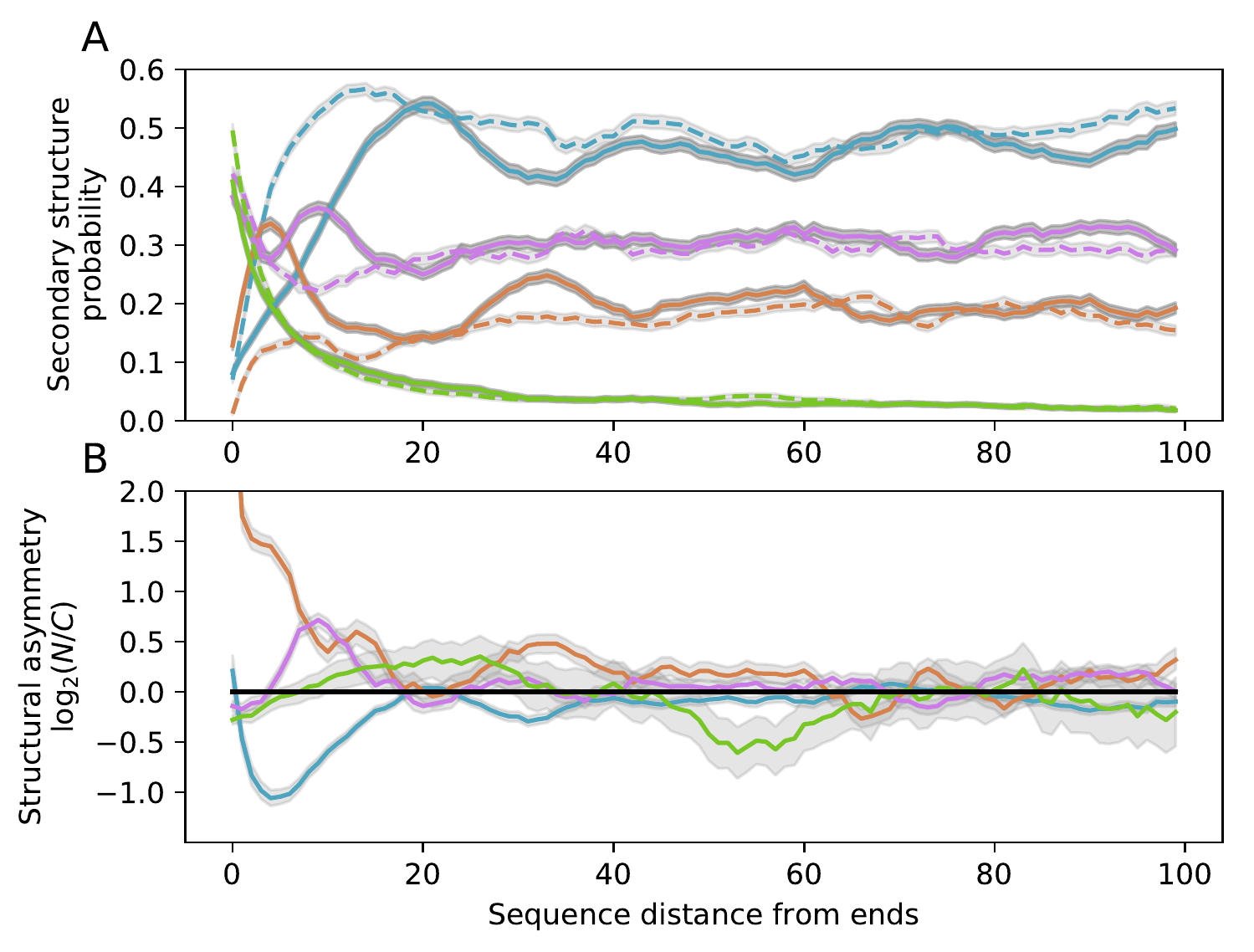}
  \caption{\label{fig:si15}
  (A) Distribution of secondary structure along the sequence as a function
  of distance from the N- and C-terminus, and (B) the structural asymmetry
  -- the ratio of the N and C distributions (in $\log_2$ scale; $\pm1$ are
  2:1 and 1:2 N/C ratios) -- for \num{4428} proteins with $179 \geq L \geq 416$
  and $21 \geq \textrm{CO} \geq 35$.
  Shading indicates bootstrapped \num{95}$\%$ confidence intervals (A-B).\\
  The asymmetry is greatest at the ends (\num{124}$\%$ more $\beta$-strands
  at the 20 residues closest to the N terminus; \num{39}$\%$ more $\alpha$-helices
  at the C terminus), and persists towards the middle (at residues 80-100, there
  are \num{7}$\%$ more $\beta$-strands at the N terminus, and \num{9}$\%$ more
  $\alpha$-helices at the C terminus.
  }
  \end{figure}

  \clearpage

  \begin{figure}[t!]  \centering
  \includegraphics[width=0.95\linewidth]{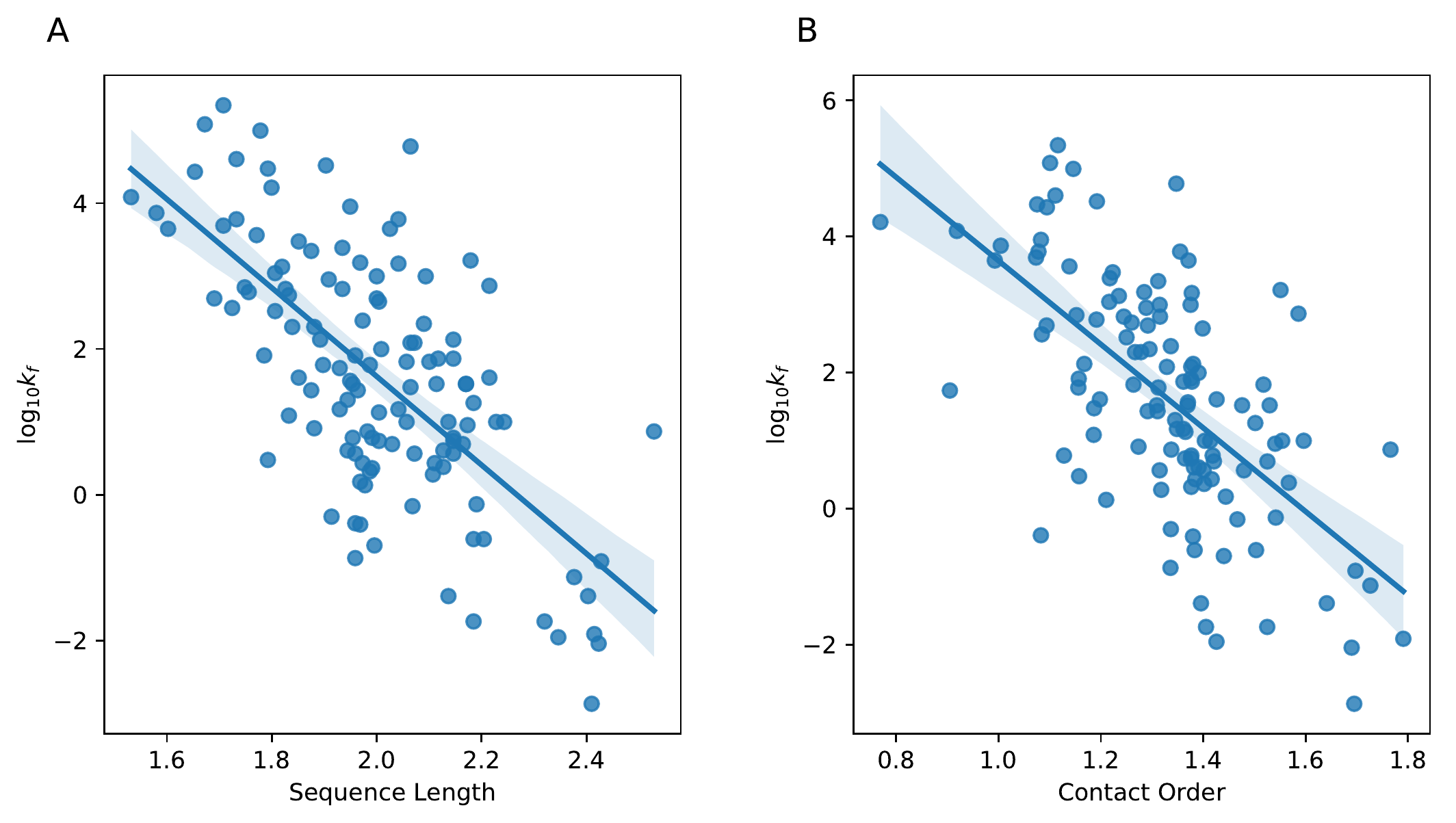}
  \caption{\label{fig:si2}
  Correlations between sequence length (\A), contact order (\B)
  and empirically-determined folding rate $k_f$ (s$^{-1}$) \cite{mansr19}.
  Two tailed Pearson's correlation gives:
  sequence length, $r=-0.68$, $p<0.005$;
  contact order, $r=-0.71$, $p<0.005$.
  }
  \end{figure}

  \clearpage

  \begin{figure}[t!]  \centering
  \includegraphics[width=0.95\linewidth]{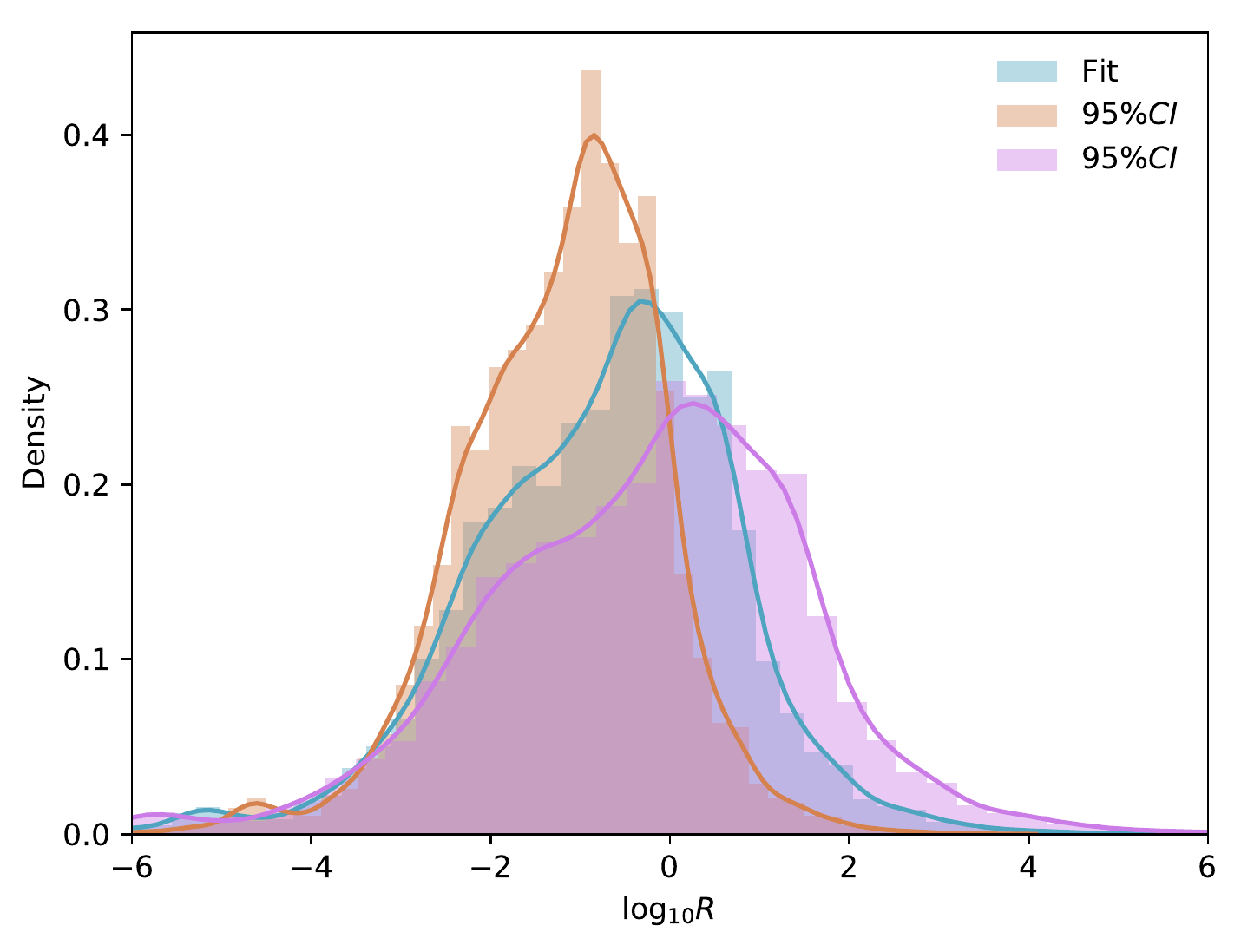}
  \caption{\label{fig:si3}
  Distribution of $R$, the ratio of folding to translation time, estimated by
  using the main fit to Eq. 7 (main text), and the $95\%$ confidence intervals
  to the fit.
  }
  \end{figure}

  \clearpage

  \begin{figure}[t!]  \centering
  \includegraphics[width=0.95\linewidth]{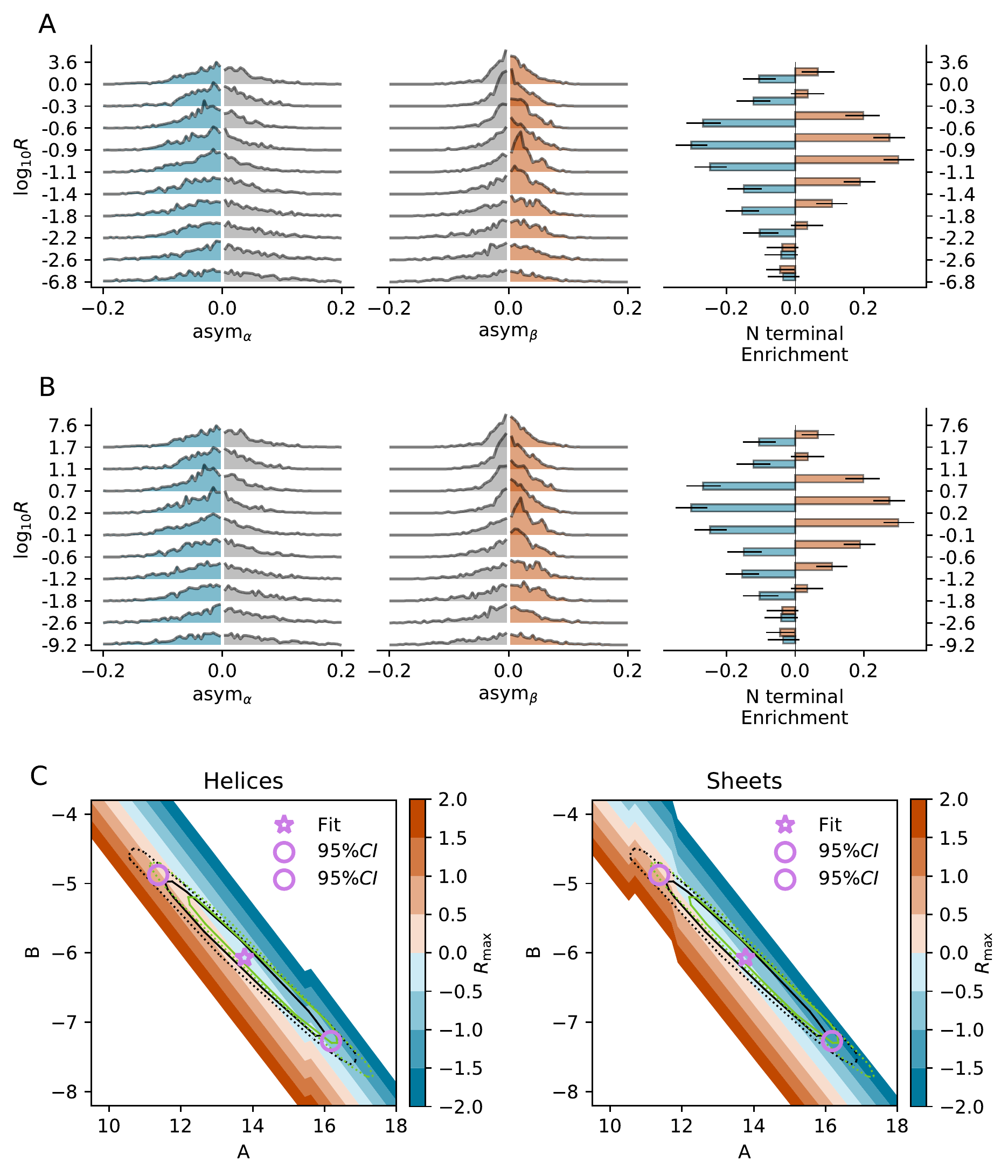}
  \caption{\label{fig:si4}
  \A-\B: $\ab$~asymmetry distributions and N terminal enrichment for the full sample according
  to the $95\%$ confidence intervals on the empirical fit for estimating $\ktrans$.
  Proteins are divided into deciles according to $R$; bin edges are shown on the y-axis.
  \C: The median of the $R$ decile that exhibits maximum asymmetry, $R_{\textrm{max}}$, as a function of
  fitting parameters $A$ and $B$. The pink star corresponds to the fit used in the main text.
  The pink circles correspond to the fits used in \A~and \B.
  The rings show the bootstrapped $95\%$ confidence interval for values of $A$ and $B$.
  Black rings are for the PFDB data set used in the main figures; green rings are for
  the less conservative PFDB data set. Solid rings indicate bootstrapped
  samples with the same sample size as the original sample; dotted rings indicate bootstrapped
  samples with half the sample size of the original sample.
  }
  \end{figure}

  \clearpage

  \begin{figure}[t!]  \centering
  \includegraphics[width=0.95\linewidth]{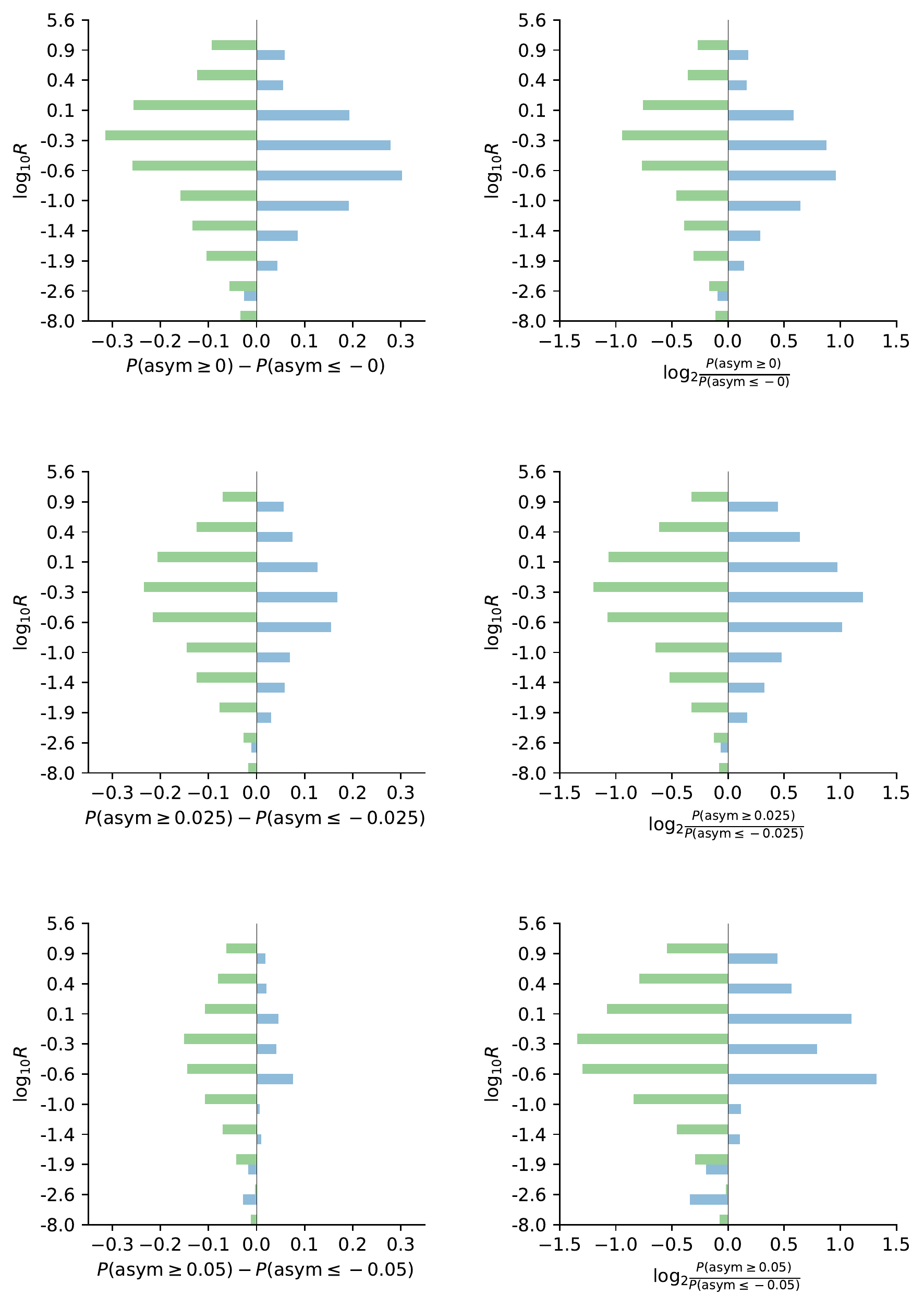}
  \caption{\label{fig:si12}
  To get a deeper understanding of what this asymmetry means,
  we consider two measures that are similar to the N terminal enrichment reported
  in the main text:
  (i) the difference in the fraction of proteins with asymmetry greater
  than $x$, and lower than $-x$ (left);
  (ii) the ratio of the fraction of proteins with asymmetry greater than $x$,
  and lower than $-x$ (right).
  We can interpret these results by considering that $L \times \textrm{asym}$
  gives us the number of extra $\beta$-strand or $\alpha$-helix residues
  at the C or the N terminus.
  Proteins are divided into deciles according to $R$;
  bin edges are shown on the y-axis.
  }
  \end{figure}

  \clearpage

  \begin{figure}[t!]  \centering
  \includegraphics[width=0.95\linewidth]{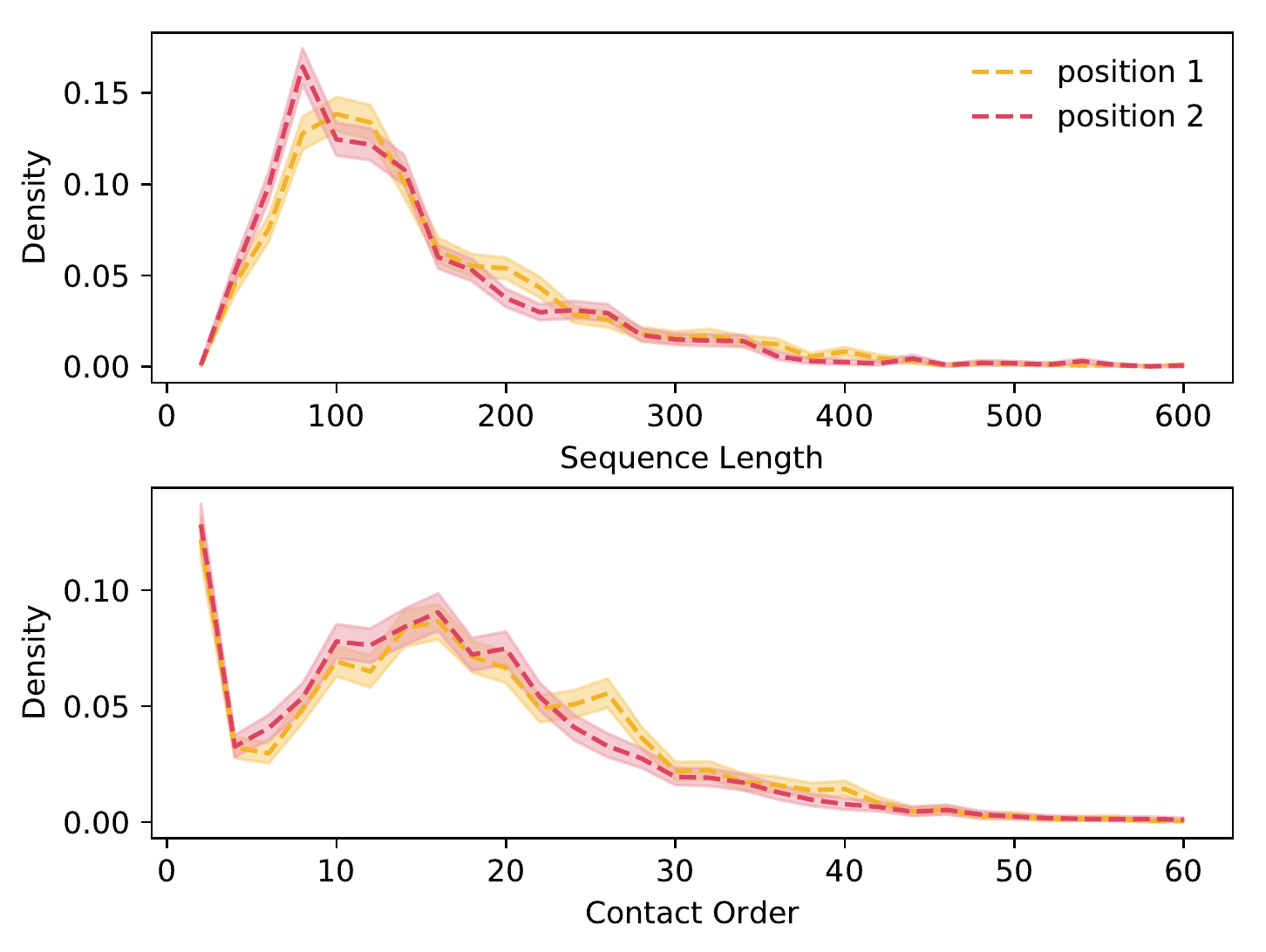}
  \caption{\label{fig:si5}
  Sequence length (top) and contact order (bottom) distributions for domains in position 1
  (near the N terminal) and position 2 in two-domain proteins.
  Shaded area indicates bootstrapped $95\%$ confidence intervals. \\ The distributions
  do not differ based on domain position, which suggests that folding time
  does not depend on domain position.
  }
  \end{figure}

  \clearpage

  \begin{figure}[t!]  \centering
  \includegraphics[width=0.95\linewidth]{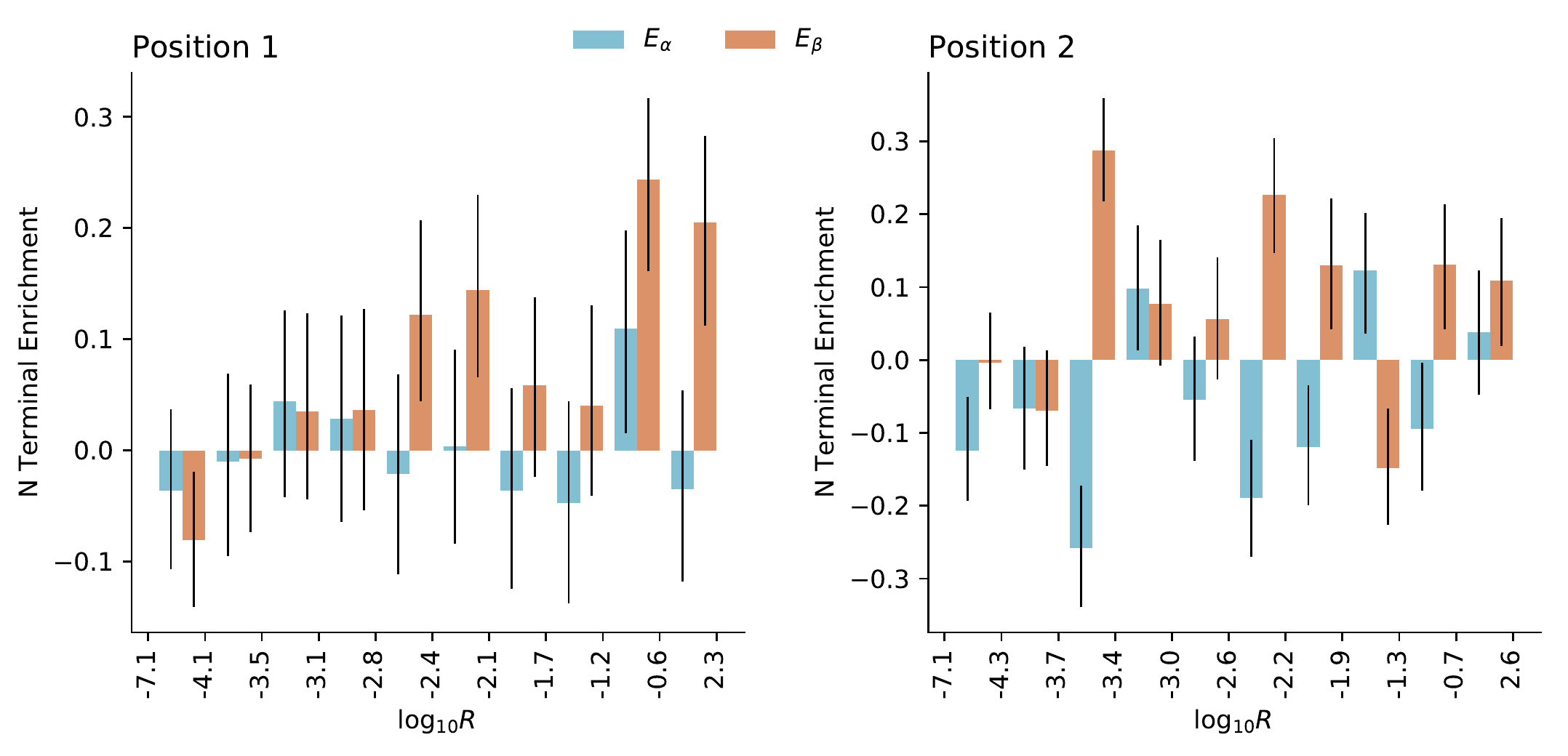}
  \caption{\label{fig:si6}
  N terminal enrichment of $\alpha$ helices ($E_{\alpha}$) and
  $\beta$ sheets ($E_{\beta}$) for individual domains within
  two-domain proteins for domains in position 1 (left; closest to the
  N terminal) and position 2 (right). Proteins are divided into deciles according to $R$;
  bin edges are shown on the x-axis; whiskers indicate bootstrapped $95\%$ confidence intervals. \\
  The data shows that domains in position 1 exhibit maximum $\beta$ asymmetry
  when $-1.2 < R < 2.4$, in agreement with the proposed `slowest-first' scheme.
  However, the confidence intervals are almost as large as the effect size, so
  we lack sufficient data to show a significant effect. Domains in position 2
  do not appear to exhibit asymmetry in agreement with the `slowest-first' scheme.
  }
  \end{figure}

  \clearpage

  \begin{figure}[t!]  \centering
  \includegraphics[width=0.95\linewidth]{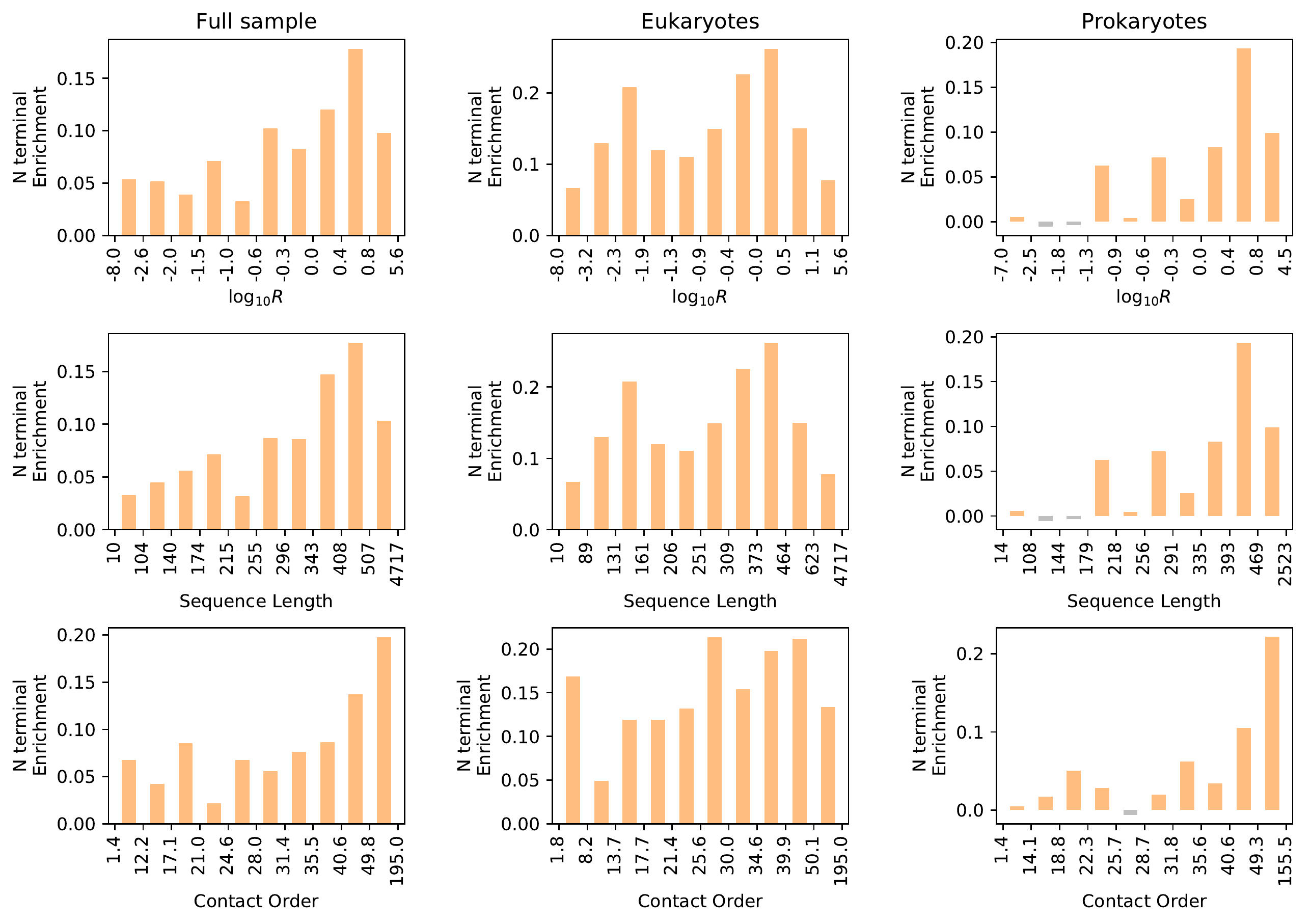}
  \caption{\label{fig:si7}
  N terminal enrichment of disordered residues as a function of
  $\log_{10}R$, Sequence Length, and Contact Order; data is shown
  for the full sample (left), eukaryotic proteins (middle), and prokaryotic
  proteins (right). Proteins are divided into deciles according to $R$ (top),
  Sequence Length (middle), and Contact Order (bottom);
  bin edges are shown on the x-axis. \\
  There is a clear association between N terminal enrichment and slow-folding
  proteins in prokaryotes, but not in eukaryotes.
  }
  \end{figure}

  \clearpage

  \clearpage

  \begin{figure}[t!]  \centering
  \includegraphics[width=0.95\linewidth]{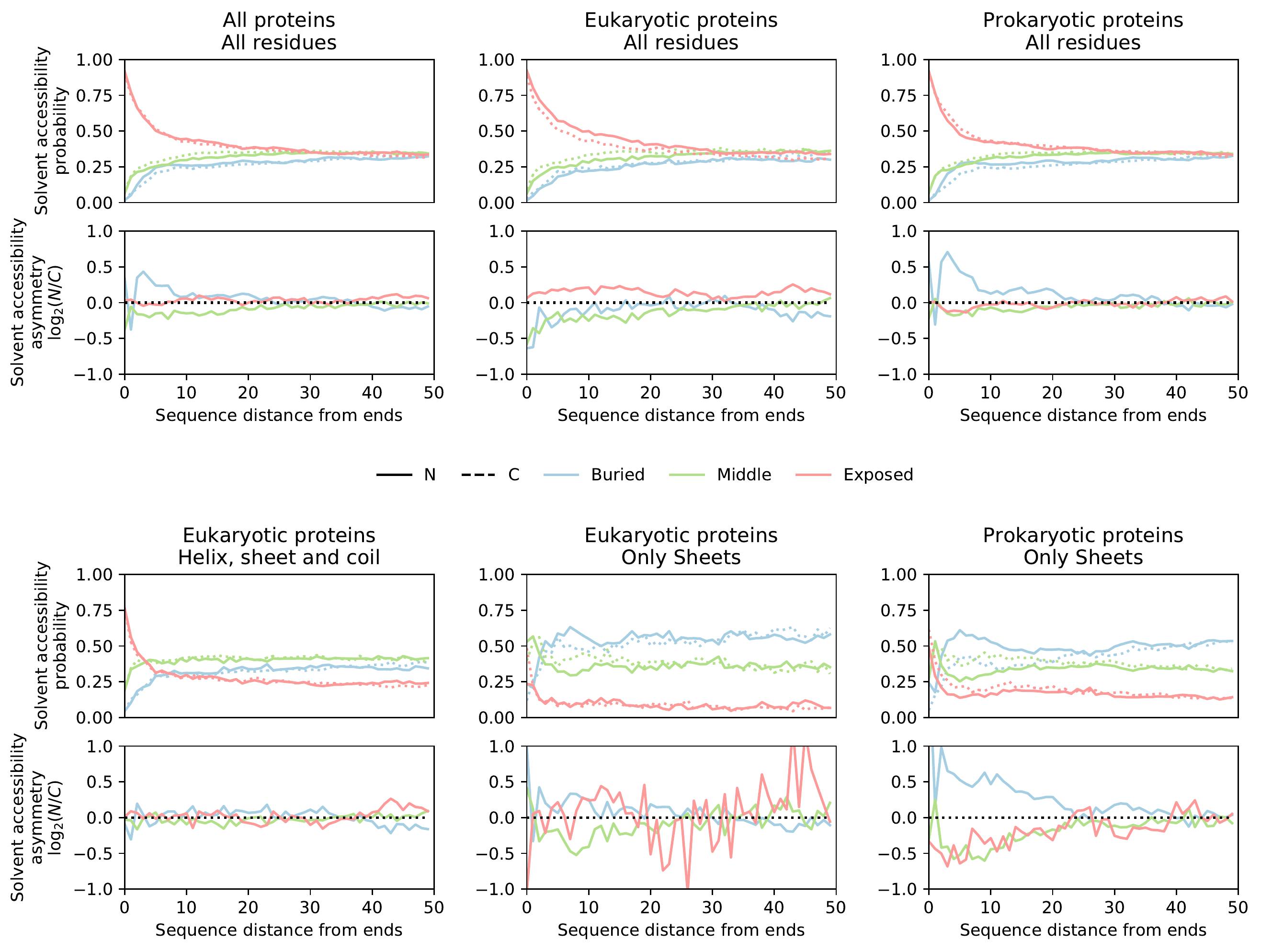}
  \caption{\label{fig:si9}
  Relative solvent accessibility (SA) and SA asymmetry as a
  function of sequence distance from the terminal residues.
  SA is calculated using the freesasa C library \cite{leejm71,mitfr16}, and residues
  are divided equally among three categories: Buried, Middle, or Exposed.
  It is impossible to calculate SA for disordered residues, and we assume that these
  are Exposed \cite{marjm13}.
  (Top) Data is shown for all proteins (left), eukayotic proteins (middle) and
  prokaryotic proteins (right).
  (Bottom) Data is shown for helix, sheet and coil residues from eukaryotic proteins (left),
  for sheet residues from eukarotic proteins (middle), and for sheet residues from
  prokaryotic proteins (right). \\
  For eukaryotic proteins overall, the N terminal has a tendency to be more exposed than
  the C terminal. However, when only considering helices, sheets and coils there is
  no significant asymmetry -- \ie~the bias for being exposed at the N terminal
  is explained by the bias for being disordered at the N terminal.
  When only considering sheets, the N terminal
  is slightly more buried, while the C terminal is more likely to contain 'Middle'
  values of SA.
  For prokaryotic proteins overall, the N terminal is more likely to buried compared
  to the C terminal, which is mainly due to $\beta$ sheets.
  }
  \end{figure}

  \clearpage

  \begin{figure}[t!]  \centering
  \includegraphics[width=0.95\linewidth]{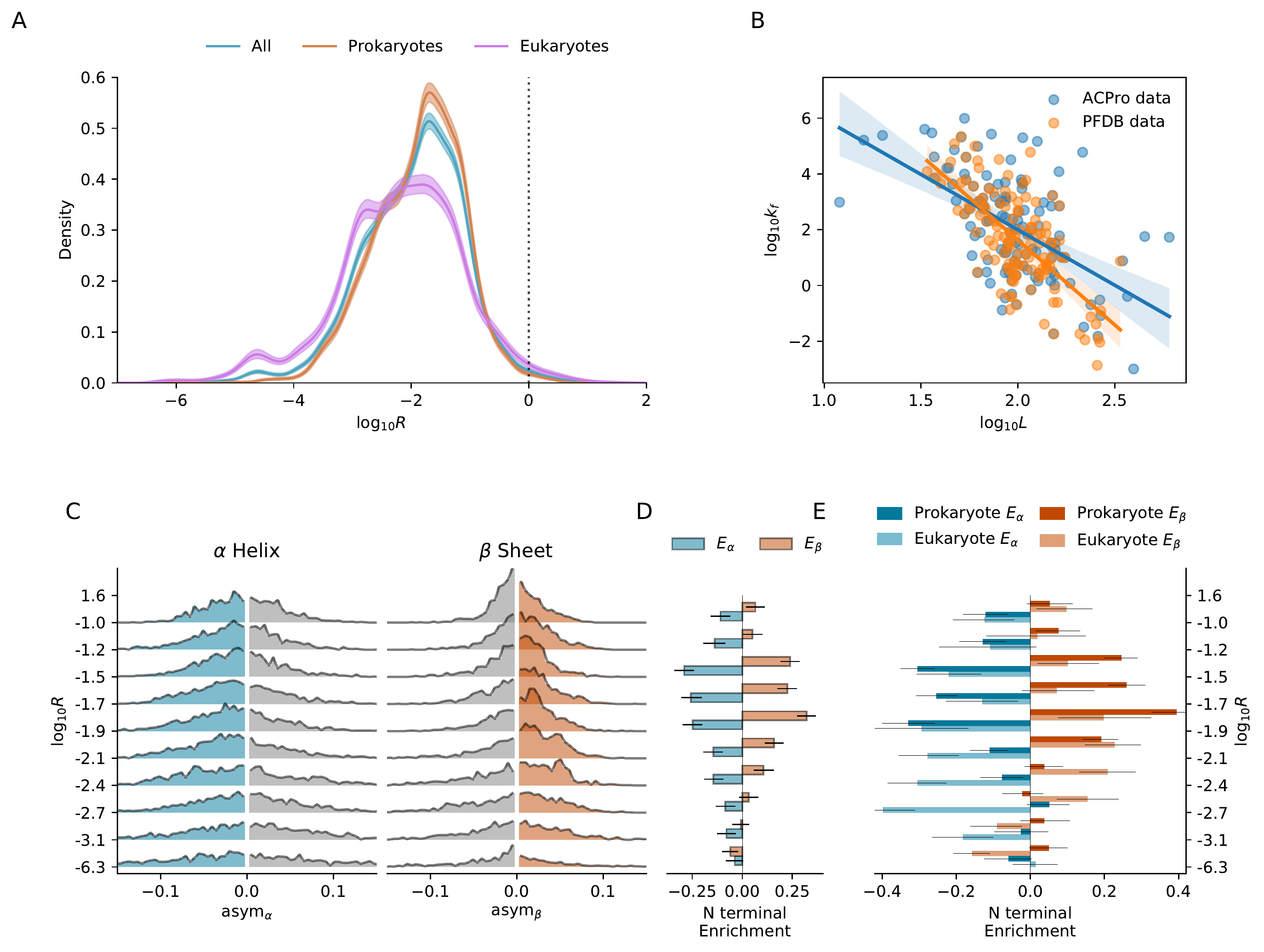}
  \caption{\label{fig:si10}
  Repeat of main text Figures 2-3 after calculating $R$, the folding/translation time ratio,
  using the ACPro database \cite{wagps14} instead of the PFDB \cite{mansr19}.
  \A: $R$ distribution for the full PDB sample, prokaryotic proteins, and eukaryotic proteins.
  \B: Scatter plot and correlation between sequence length $L$ and folding rate $k_f$ (s$^{-1}$) for
  both the ACPro and PFDB databases. Shaded area indicates $95\%$ confidence interval
  of the linear fit.
  \C: $\ab$~asymmetry distributions as a function of $R$.
  Proteins are divided into deciles according to $R$;
  bin edges are shown on the y-axis.
  \D: N terminal enrichment -- the degree to which sheets/helices
  are enriched in the N over the C terminal -- 
  is shown for the deciles given in C.
  \E: N terminal enrichment as a function of $R$ for \num{4633} eukaryotic proteins
  and \num{10400} prokaryotic proteins. Proteins are divided into bins according to $R$;
  bin edges, shown on the y-axis, are the same as in C-D.
  Whiskers indicate $95\%$ confidence intervals. \\
  In contrast with the results obtained using the PFDB, estimating $R$ using the ACPro
  database results in the prediction that for most proteins translation time
  is considerably longer than folding time. The principal reason for this appears to be
  a few proteins with either few residues ($<34$) or many residues ($>300$) that are present
  in the ACPro data but not the PFDB. These residues change the gradient of the fit
  such that large proteins are predicted to fold much faster according to the ACPro
  fit. \citet{mansr19} cite several reasons for why certain entries in ACPro were
  not used in the PFDB. Even despite these differences, the region of maximal
  asymmetry is predicted to be $-1.9 < R < -1.2$, where non-negligible acceleration
  of CTF is still possible.
  }
  \end{figure}

  \clearpage

  \begin{figure}[t!]  \centering
  \includegraphics[width=0.70\linewidth]{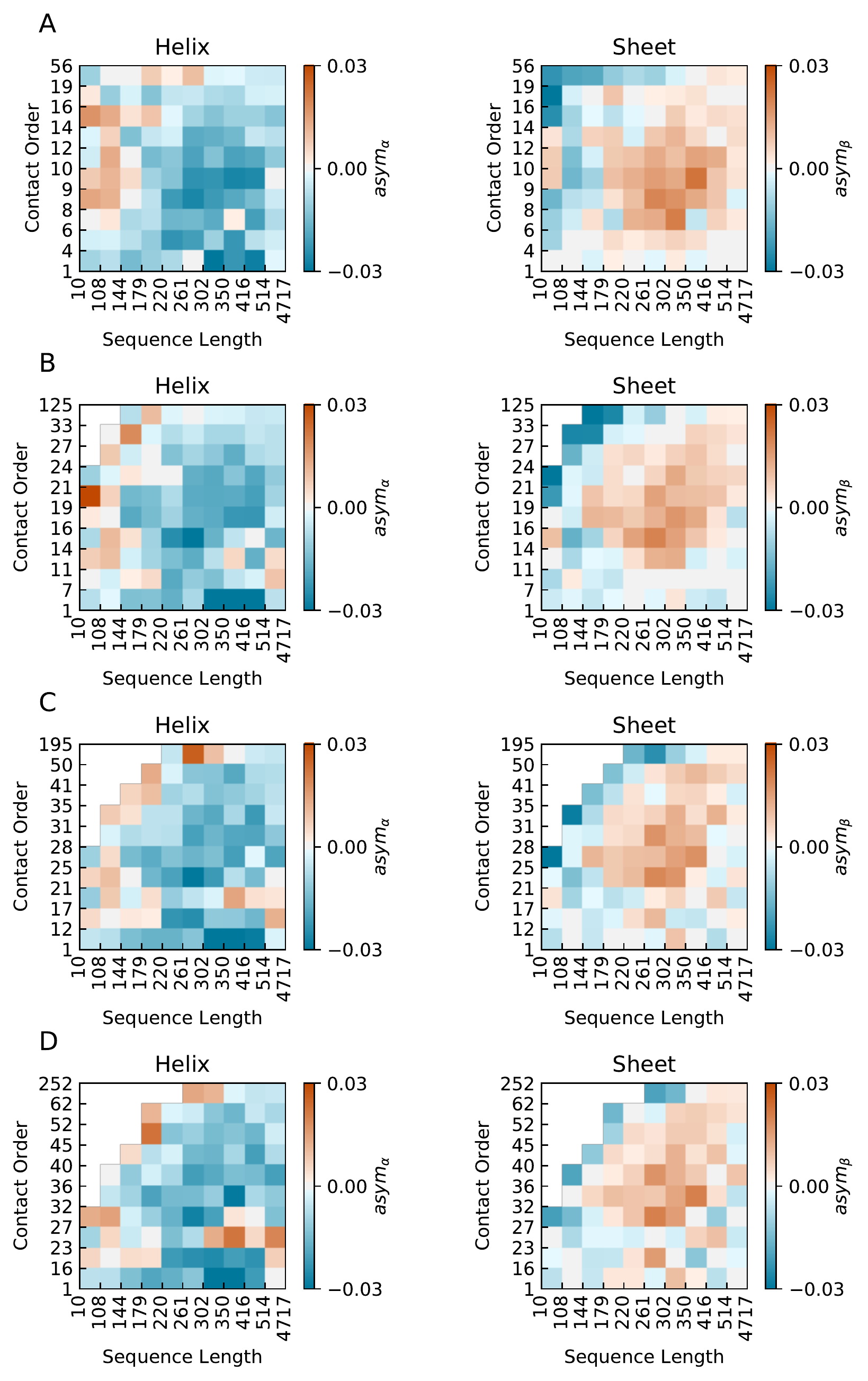}
  \caption{\label{fig:si11}
  Correlation between sequence length, contact order and $\ab$~asymmetry.
  Separate plots are shown for different values of the cutoff used to
  calculate contact order: \A, $6$ \AA; \B, $8$ \AA; \C, $10$ \AA; \D, $12$ \AA.
  }
  \end{figure}

  \clearpage

\bibliographystyle{unsrtnat}
\bibliography{supporting}